\documentclass{IEEEtran}
\usepackage[utf8]{inputenc}

\usepackage{amsmath}
\usepackage{bm}
\usepackage{graphicx}
\usepackage{placeins}

\everypar{\looseness=-1}

\title{Design of Spatial-Spectral Filters for CT\\ Material Decomposition\vspace{-2mm}}
\author{Matthew Tivnan, Wenying Wang, Grace Gang, J. Webster Stayman\vspace{-10mm}}
\date{August 2019}
\begin{document}

\maketitle

\begin{abstract}
Spectral CT has shown promise for high-sensitivity quantitative imaging and material decomposition. This work presents a new device called a spatial-spectral filter (SSF) which consists of a tiled array of filter materials positioned near the x-ray source that is used to modulate the spectral shape of the x-ray beam. The filter is moved to obtain projection data that is sparse in each spectral channel. To process this sparse data, we employ a direct model-based material decomposition (MBMD) to reconstruct basis material density images directly from the SSF CT data. To evaluate different possible SSF designs, we define a new Fisher-information-based predictive image quality metric called separability index which characterizes the ability of a spectral CT system to distinguish between the signals from two or more materials. This predictive metric is used to define a system design optimization framework. We have applied this framework to find optimized combinations of filter materials, filter tile widths, and source settings for SSF CT. We conducted simulation-based design optimization study and separability-optimized filter designs are presented for water/iodine imaging and water/iodine/gadolinium/gold imaging for different patient sizes. Finally, we present MBMD results using simulated SSF CT data using the optimized designs to demonstrate the ability to reconstruct basis material density images and to show the benefits of the optimized designs.
\end{abstract}

\vspace{-6mm}

\section{Introduction}

% Mitch goodsitt houndsfield unit variability. patient size.

% rotating filter. 

% spectral modulator adam wang

% AAPM novel CT hardware.

% Ge Wang GOLF

% beam blocker

% slow kv switching

High-sensitivity material discrimination is an important goal for the next generation of x-ray computed tomography systems. One way to accomplish this goal is by incorporating varied photon energy spectral sensitivities into one CT acquisition. Even if two materials map to the same overall attenuation (as measured by Hounsfield units in conventional CT) they can be distinguished by the relative response across multiple sensitivity channels on a spectral CT system. Furthermore, if a finite number of basis materials can be assumed, the material densities can be estimated directly via a material decomposition algorithm.

Material density estimation is an important development that moves CT into the domain of true quantitative imaging. Recent years have seen an explosion of medical image post-processing such as anatomical segmentation, the extraction of radiomic features and computer-automated diagnosis. These image processing software tools often implicitly assume that the input images represent the same physical quantity with the same units. However, it is well known that Hounsfield units vary based on the source settings, detector properties and other factors. Therefore, there is a need for standardization of the physical meaning of the numeric values which make up CT images regardless of the imaging hardware or data processing software used to generate them. Material density images serve this purpose because they have a clear physical meaning that can provide a unified quantitative basis across various systems.

Spectral CT and material density estimation also enable imaging studies that would not have been possible with conventional CT. For example, decomposition into water and calcium allows for quantitative measures of bone density. Decomposition of iodine contrast agent, enables virtual non-contrast imaging without the need for a digital subtraction step. Three or more spectral sensitivity channels would enable multi-contrast-enhanced imaging studies. For example, protocols have already been developed for multiphasic liver imaging using two different contrast agents (e.g. iodine and gadolinium) with time-delayed injections to acquire the arterial enhancement phase, venous enhancement phase, and non-contrast phase in a single acquisition \cite{muenzel2017simultaneous}.

Spectral CT encompasses a class of technologies which modulate the x-ray energy sensitivity spectra used to sample the energy-dependent attenuation properties of the patient. One strategy is to incorporate varied spectral sensitivities using energy-sensitive detectors. Some existing dual-energy CT systems are enabled by dual-layer detectors \cite{rassouli2017detector}, where the low-energy photons are preferentially absorbed in the first layer, which corresponds to the low-energy channel. The high-energy channel is the second layer which interacts with photons which have passed through the first layer. Novel photon-counting detectors are promising for the future of spectral CT because they have the ability to directly discriminate the energy of detected photons \cite{leng2019photon}. However, currently the vast majority of existing CT systems use conventional energy-integrating detectors. Some spectral CT technologies are compatible with energy-integrating detectors and are therefore candidates for incremental modification of existing imaging systems. These include dual-source CT \cite{flohr2006first}, where there are two physical sources, each tuned to a different peak tube voltage as well as kV-switching \cite{zou2008analysis} strategies where there the source settings are varied as a function of view angle. 

Another strategy for spectral CT with energy integrating detectors is to modulate the spectrum of the incident x-ray beam using filtration materials, particularly those with a k-edges at useful energy levels. Split-filter designs \cite{rutt1980split} have been implemented using materials such as tin positioned near the source to filter a portion of the beam that covers half of the fan angles. For a circular source trajectory, one can take advantage of the data redundancy of 360 degree acquisitions to produce two complete CT datasets with two different spectral sensitivities from one acquisition using a split-filter.

In this paper, a novel spectral modulator is presented called a spatial-spectral filter (SSF) which uses tiled source-side filters to enable an arbitrary number of spectral channels. 
These filters present possible advantages in designing varied and customized spectral permitting optimization of spectral sensitivity for specific tasks. Design of these filters presents both flexibility and challenges due to the trade-off between spatial and spectral sampling.

There are many possible filter layouts, but in general, a particular spectral channel associated with one filter material is sparse and does not constitute a complete CT dataset in terms of spatial sampling of the projection domain. Therefore, alongside our presentation of this new device, we provide the data processing strategies needed to perform spatial reconstruction and material decomposition to estimate material density images directly from sparse spectral CT data acquired with a SSF.

To consider optimal designs under different situations, we introduce a new predictive image quality metric called separability index which characterizes the ability of a spectral CT system to distinguish between the signals from two or more materials. This metric is related to the Fisher information matrix and is intended to quantify the advantage a spectral CT system would have in discriminating stimuli over a traditional single-energy CT device.

% The main advantages of SSFs are patient-dose-efficient spectral diversity and a flexible design that can be tuned to optimize performance for a given task. The main challenge is an irregular spatial spectral sampling pattern.

In previous preliminary studies, we have demonstrated that full material decomposition is possible using SSF CT data despite the sparsity of each channel in the projection domain \cite{tivnan2019physical}. We have also presented preliminary results characterizing the impact of various design parameters on the material density image quality \cite{tivnan2019optimized} \cite{tivnan2019designing}. In this work, a detailed physical description of the SSF is presented including an enumeration of the tuneable design parameters. We provide an advanced physical model for polyenergetic x-ray sources, filtration, attenuation, and detection, including models for non-ideal effects such as spectral blur and quantum noise. An iterative model-based material decomposition algorithm is also introduced including regularization strategies to allow for spatial reconstruction from incomplete CT data. In conjunction with the new separability metric, optimized designs are sought for different imaging conditions. Finally, simulation studies are conducted for a variety of SSF designs to demonstrate improved performance for designs that have been optimized for separability.

\begin{figure*}[htb!]
    \centering
    \includegraphics[trim={135mm 0mm 0mm 0mm } , clip, width=\textwidth ] {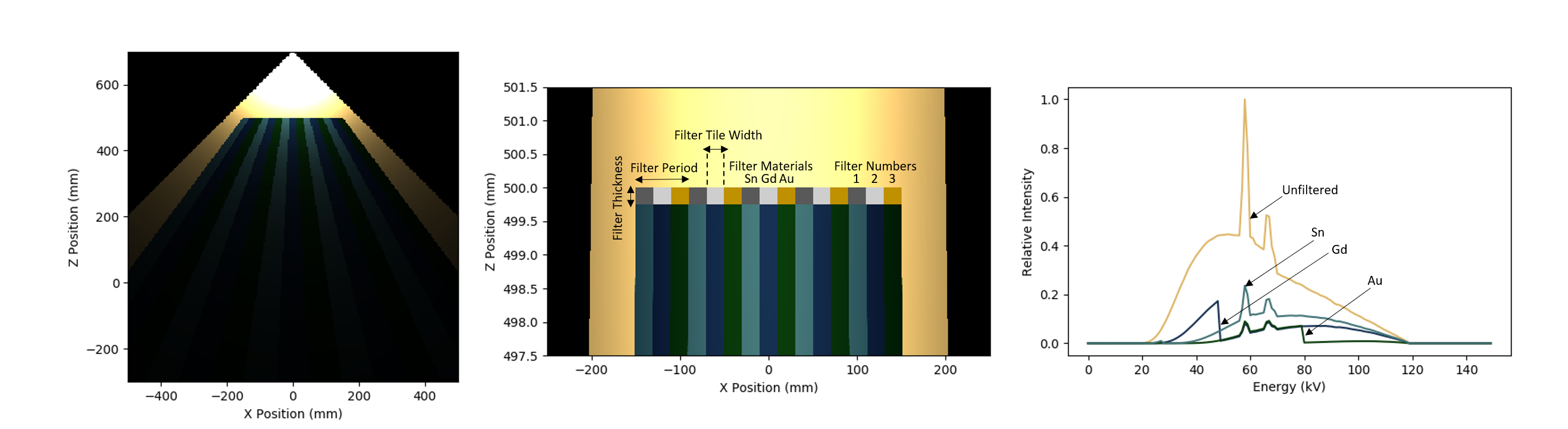}
    \vspace*{-.5in}
    \caption{SSF using k-edge materials (Sn, Gd, Au). A uniform polyenergetic beam generated by a 120kvP source is incident on the SSF (from the top of figure) which divides it into spectrally varied beamlets (bottom half of figure). The spectra on the right-hand plot show discrete drops at the k-edge of Gd (50.24 keV) and the k-edge of Au (80.72).}
    \label{fig:ssfCartoonA}
    \vspace{-6mm}
\end{figure*}

% \begin{figure*}[ht]
%     \centering
%     \includegraphics[trim={135mm 0mm 0mm 0mm } , clip, width=\textwidth ]{figures/ssfCartoon_B_combined.png}
    % \caption{SSF using non-k-edge materials (Ni, Cu, Zn). The materials are very similar in terms of x-ray filtration so the resulting spectra are almost the same even though there are three different filters.}
%     \label{fig:ssfCartoonB}
% \end{figure*}

\section{Methods}

\subsection{Spatial Spectral Filters}

The beam from a conventional x-ray tube consists of x-ray photons with varied energies less than or equal to the source voltage times the charge of an electron. This spectrum is distributed according to idealized bremsstrahlung with additional filtration from the tungsten anode and other source components in the path of the beam. The proposed device is a patterned filter positioned between the x-ray source and the patient. The filter consists of a tiled array different materials (e.g., metallic foils). The polyenergetic beam from the source is incident on the filter. X-ray photons are attenuated according to the unique energy-dependent attenuation associated with the filter material it passes through at each position. Different paths of x-ray propagation (i.e., rays) will intersect different parts of the tiled filter and therefore be filtered by different materials. In this way, the ordinarily spatially uniform polyenergetic incident beam is divided into spectrally varied beamlets. The SSF modulates the spatial-spectral distribution of photons incident on the patient and consequently, the spatially-dependent sensitivity spectra of the lines of projection measured by a detector on the opposite side of the patient. A diagram showing the SSF is provided in Figure \ref{fig:ssfCartoonA}.

The materials and thicknesses and spatial distribution of the filters can be chosen to produce the desired shape of the sensitivity spectra. One important design consideration for metallic filters is the position of the k-edge. As a general trend, materials tend to attenuate low-energy photons at a greater rate than high-energy photons. However, there are discrete positive jumps in attenuation at the k-shell energy for a given element since another quantized level of photoelectric absorption is available to photons above this energy level. Materials that are useful have a k-edge in the diagonostic energy range. For example, tin (atomic number 50) has a k-edge 29.20 keV and bismuth (atomic number 84) has a k-edge at 90.53 keV. Materials with an atomic number greater than 84, and a corresponding higher k-shell energy, are typically unstable, radioactive, or toxic. Using varied k-edge materials leads to a more diverse set of sensitivity spectra. Spectral diversity is an important consideration for separability in joint material decomposition and spatial reconstruction problems. An example of a SSF using materials (Sn, Gd, Au) with k-shell energies in the diagnostic range is shown in Figure \ref{fig:ssfCartoonA}.

% and an example of a SSF using materials (Ni, Cu, Zn) with k-shell energies outside the diagnostic range is shown in Figure \ref{fig:ssfCartoonB}. 

To improve the spatial-spectral sampling pattern, the SSF position is allowed to translate as a function of view angle. For example, if the SSF is not translated, then the portion of an object positioned at the axis of rotation will only be sampled by one spectrum. This is generally not sufficient for material decomposition.

There are many SSF system design parameters which control spatial-spectral sampling. In this work, we constrain the SSF to a repeating 1-dimensional array of rectangular filter tiles as shown in Figure \ref{fig:ssfCartoonA}. Within these constraints, the remaining design parameters which control the static physical design of the SSF include:

\begin{enumerate}
    \item the thickness of the filter tiles
    \item the period/length of the repeating pattern of filter tiles
    \item the number of filter tiles per period
    \item the material of each filter tile
    \item the ordering of materials in the repeating pattern
    \item the relative width of each filter tile (must sum to length period)
\end{enumerate}

Additionally, there are some parameters which do not pertain to the construction of the SSF. These have an effect on the system spatial-spectral sensitivity and sampling but can be adjusted between or during scans. These adjustable design parameters for SSF CT include:

\begin{enumerate}
    \item the trajectory of the SSF translation
    \item the trajectory of the source and detector
    \item the distances between the source, the SSF, the object, and the detector
    \item the x-ray tube voltage
    \item the x-ray tube current
\end{enumerate}

Adjusting these parameters allows SSF to be used to enable adaptive spectral CT acquisitions. By controlling the trajectory of the SSF, we can control which parts of the object are sampled by each sensitivity spectra. The x-ray tube current and voltage can also be modulated for further control over the number and spectral distribution of photons incident on the patient. In this work we focus on optimization of the intrinsic filter parameters except for ordering, which was previously found not to be a significant factor \cite{tivnan2019optimized}. Similarly, the CT system design was fixed with a circular cone-beam orbit and geometry; while controllable factors like tube voltage are optimized. 
%Translation? fixed? (not that important based on studies?)
%%%% WEB %%%% Translation? (You had the above uncommented)
%% MTT - resolved. For the specific studies, I described the filter motion pattern. I think its fine to leave as a hypothetical here.

As compared to some other spectral CT technologies, the data collected using SSFs is  incomplete in that it cannot be divided into a full CT dataset for each spectral ``channel''. That is, if each filter material is considered as a separate spectral channel, then there is a sparse CT projection-domain dataset collected for each channel. Importantly, the spatial sampling geometry is still complete if all spectral channels are taken together. Further, if each part of the object is exposed to a different sensitivity spectra as a function of view, we have spectral information that can be used for material decomposition. Such strategies involving trade-offs between spatial and spectral sampling have been widely explored in other imaging systems (e.g. spectral imaging from air- and space-borne cameras \cite{wang2013spatial}.)

In the next section we present advanced data processing strategies to handle this data and strategies for optimizing the design. This includes a statistical estimation algorithm for jointly solving the spatial reconstruction and material decomposition problems as well as regularization schemes to handle the sparse spatial-spectral sampling. Moreover, a new performance metric of material separability is introduced to quantitatively predict and select optimized filter designs.

\vspace{-3mm}

\subsection{Physical Models and Material Decomposition}

% NOTE:
% Cite some papers using model in EQ 1

In this section, we introduce a physical model for polyenergetic x-ray transmission and detection in spectral CT data acquisitions which can be used to model SSF CT systems.

\subsubsection{Spectral CT Measurement Model} Given a multi-material density image, $\mathbf{x}$ (all voxels and materials represented as a single column vector), a discretized spectral CT measurement model, $\mathbf{\bar{y}}(\mathbf{x})$, may be written as 

\vspace{-3mm}

\begin{equation}
    \mathbf{\bar{y}}(\mathbf{x}) = \mathbf{S} \exp(-\mathbf{Q}\mathbf{A}\mathbf{x})
\end{equation}

\vspace{-2mm}

\noindent where $\mathbf{A}$ is a forward-projection matrix operator which models line integrals, $\mathbf{Q}$ contains the mass attenuation coefficients for each basis material, and $\mathbf{S}$ is the projection- and energy-dependent spectral sensitivity. This matrix formulation has previously been used in \cite{tilley2018model} for model-based material decomposition but also represents the same basic form as models used in \cite{barber2016algorithm}.
The shapes of these matrix operators is given in Table \ref{tab:matrixShapes}.

Note, that $\mathbf{A}$ is block-diagonal across basis materials (requiring one forward projection per material) and $\mathbf{Q}$ is block-diagonal across projections. However, the system spectral sensitivity $\mathbf{S}$ can depend on both photon energy, projection, and view angle, as is the case for SSF CT which modulates the spatial-spectral 0sampling distribution differently for each projection and view.

\subsubsection{Noise Model} This work focuses on indirect energy-integrating detectors, and so we expand $\mathbf{S}$ as follows:

\vspace{-3mm}

\begin{equation}
    \mathbf{\bar{y}}(\mathbf{x}) = \mathbf{G}_4 \mathbf{B}_3 \mathbf{S}_2 \mathbf{S}_1 \mathbf{S}_0 \exp(-\mathbf{Q}\mathbf{A}\mathbf{x})
\end{equation}

\vspace{-2mm}

\noindent where the numerical subscripts loosely follow the stages of the signal and noise transfer models for detectors as presented in \cite{siewerdsen1997empirical}. The operator $\mathbf{S}_0$ models the energy-dependent mean number of incident x-ray photons (after SSF filtration) for each line of response, $\mathbf{S}_1$ models the energy-dependent probability of interaction with the scintillator, $\mathbf{S}_2$ models the energy-dependent conversion to secondary quanta (optical photons) in the scintillator, $\mathbf{B}_3$ is a scintillator blur model, and $\mathbf{G}_4$ is a diagonal scaling operator used to capture optical coupling and other overall gain effects. 

\begin{table}[]
    \label{tab:matrixShapes}
    \caption{Shapes and other properties of the matrix operators in the spectral CT measurement model}
    \centering
    \begin{tabular}{|c|c|}
        \hline Matrix Operator & Shape  \\ \hline
        $\mathbf{A}$ & $\Big(N_\text{pixels} N_\text{materials} \times N_\text{voxels} N_\text{materials}\Big)$ \\
        \hline
        $\mathbf{Q}$ & $\Big(N_\text{pixels} N_\text{energies} \times N_\text{pixels} N_\text{materials}\Big)$ \\
        \hline
        $\mathbf{S}$ & $\Big(N_\text{pixels} N_\text{channels} \times N_\text{pixels} N_\text{energies}\Big)$ \\ \hline
    \end{tabular}
    \vspace{2mm}
    \vspace{-.3in}
\end{table}

We assume that photons generated by the x-ray source for each energy level are Poisson-distributed and uncorrelated. Further, we assume that pre-filtration by the SSF, absorption by the patient, and scintillator interaction are binomial selection processes, so the quantity $\mathbf{S}_1 \mathbf{S}_0 \exp(-\mathbf{Q}\mathbf{A}\mathbf{x})$, which represents x-ray photons which are detected, is also Poisson distributed and  uncorrelated (because $\mathbf{S}_1$ is typically block diagonal across projections). However, we model $\mathbf{S}_2$, $\mathbf{B}_3$, and $\mathbf{G}_4$ as deterministic operators, so when approximating the data covariance for a given object, $\mathbf{x}$, we use the expression below.

\vspace{-5mm}

\begin{equation}
    \mathbf{\Sigma_y} =  \mathbf{G}_4 \mathbf{B}_3 \mathbf{S_2} D\{\mathbf{S}_1 \mathbf{S}_0 \exp(-\mathbf{Q}\mathbf{A}\mathbf{x})\} \mathbf{S}_2^T \mathbf{B}_3^T \mathbf{G}_4^T 
\end{equation}

\subsubsection{Model-Based Material Decomposition} 

% NOTE: Acknowledge that there are other MBMD algorithms

There are several possible approaches to MBMD which have the ability estimate the material density images directly from spectral CT data \cite{long2014multi} \cite{barber2016algorithm}. We use the SPS algorithm described in \cite{tilley2018model} which minimizes the following penalized least-squares objective function:
\vspace{-2mm}
\begin{equation}
    \Phi(\mathbf{x}, \mathbf{y}) = (\mathbf{y} - \mathbf{\bar{y}}(\mathbf{x}))^T \mathbf{\Sigma_y^{-1}} (\mathbf{y} - \mathbf{\bar{y}}(\mathbf{x})) + \mathbf{x}^T \mathbf{R} \mathbf{x}
    \label{eq:objectiveFunction}
\end{equation}
\begin{equation}
    \mathbf{\hat{x}}(\mathbf{y}) = \underset{\mathbf{x}}{\text{argmin}} \enspace \Phi(\mathbf{x}, \mathbf{y}).
\end{equation}

\vspace{-2mm}

The first term in \eqref{eq:objectiveFunction} is the data fidelity term and the second is a cross-material quadratic roughness penalty described in \cite{wang2019generalized}. That quadratic penalty is a refinement of the familiar pairwise quadratic penalty between neighboring voxel differences, except that an additional pairwise penalty between material type is applied. The weights on those penalties are chosen to minimize cross-material regularization bias as dscribed in \cite{wang2020prospective}. 
%%% WEB - see elsewhere cross-penalty weights or no?
% MTT - resolved. added citation to another of wenyings paper aboutoptimizin cross penalty weights.
\subsubsection{Dose Normalization} 

For fair comparisons between various system designs, we normalize the dose absorbed by the patient using the following formula.
\vspace{-2mm}
\begin{equation}
    \text{Dose} = (1.602 \times 10^{-13} \enspace \frac{\text{mJ}}{\text{keV}}) \enspace  \boldsymbol{\varepsilon}^T \mathbf{S}_0 (\mathbf{1} - \exp(-\mathbf{Q}\mathbf{A}\mathbf{x}))
    \label{eq:dose}
\end{equation}
\vspace{-1mm}
where $\mathbf{S}_0 (\mathbf{1} - \exp(-\mathbf{Q}\mathbf{A}\mathbf{x}))$ are the x-ray photons absorbed by the patient, and $\boldsymbol{\varepsilon}$ are the energies in keV of those photons. This presumes that the deposited energy is equal to the energy of all attenuated x-ray photons ignoring scattered photons that leave the patient.

\subsubsection{Modeling Spatial-Spectral Filters}

% NOTE: we may need to do some kind of validation. possibly experiment to show inside/outside exponential is not a large effect. maybe include as appendix.

\begin{figure}[htb!]
    \centering
    \includegraphics[width=0.45\textwidth]{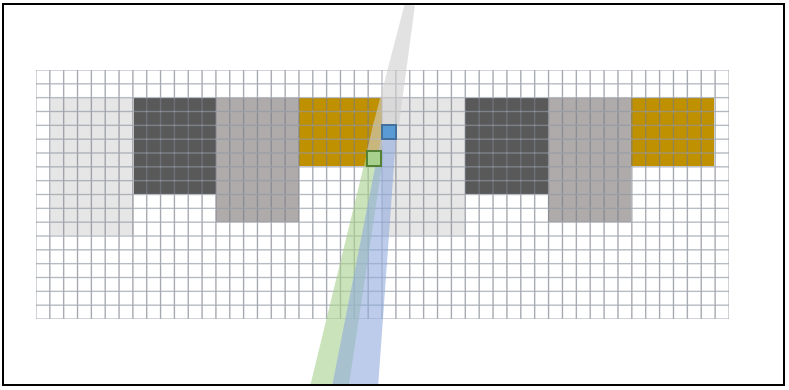}
    \vspace{-3mm}
    \caption{Forward projecting a high-resolution voxelized model for spatial-spectral filters enables models for varied geometric designs as well as spectral blur effects. Note that the diagram is not to scale. }
    \vspace{-7mm}
    \label{fig:ssfModel}
\end{figure}

Beam filtration by the SSF occurs on the source side, so it is modeled in the $\mathbf{S}_0$ term. This is beneficial as compared to detector-side filtration because the photons absorbed in the filter for the purpose of spectral shaping never actually reach the patient and therefore do not contribute to dose. The spectral modulation from an SSF is given by the following formula:
\begin{equation}
    \mathbf{S}_0 = D\{\exp(-\mathbf{Q}_\text{SSF} \mathbf{B}_\text{SSF} \mathbf{A}_\text{SSF} \mathbf{x}_\text{SSF})\} D\{\mathbf{s}_0\}.
\end{equation}
\noindent Here, $\mathbf{s}_0$ is the number of photons generated by the source at each energy before filtration, $\mathbf{x}_\text{SSF}$ is a high-resolution voxelized model of the SSF, $\mathbf{B}_\text{SSF}$ is a blur model for spectral mixing effects at beamlet boundaries (due to an extended focal spot), $\mathbf{A}_\text{SSF}$ is the system geometry, and $\mathbf{Q}_\text{SSF}$ contains the mass attenuation spectra for filter materials in the SSF as illustrated in Figure \ref{fig:ssfModel}.

Design parameters that control the physical construction of the filter are modeled in $\mathbf{x}_\text{SSF}$. This includes the filter thickness, potential gaps between filter tiles, and filter tile width among other parameters. All parameters related to the system geometry and sampling, including the source-to-filter distance, source-to-axis distance, source-to-detector distance, and filter motion trajectories are modeled in $\mathbf{A}_\text{SSF}$. Blur due to an extended focal spot as well as blur due to filter motion (for a continuous acquisition) can be modeled in $\mathbf{B}_\text{SSF}$. That is, before the $\mathbf{B}_\text{SSF}$ operation, the beamlets are perfectly formed with sharp edges associated with a point source and after the operation the beamlets are blurred (thus spectrally mixed) to model the effects of an extended focal spot.

In previous work \cite{tivnan2019designing, tivnan2019physical}, we have presented evidence that faster filter motion trajectories and narrower filter tile widths lead to a greater degree of uniformity in the spatial-spectral sampling pattern. 

\vspace{-5mm}

\subsection{Image Properties and Performance Evaluation}

\vspace{-1mm}

\begin{figure*}[htb!]
    \centering
    \includegraphics[width=\textwidth]{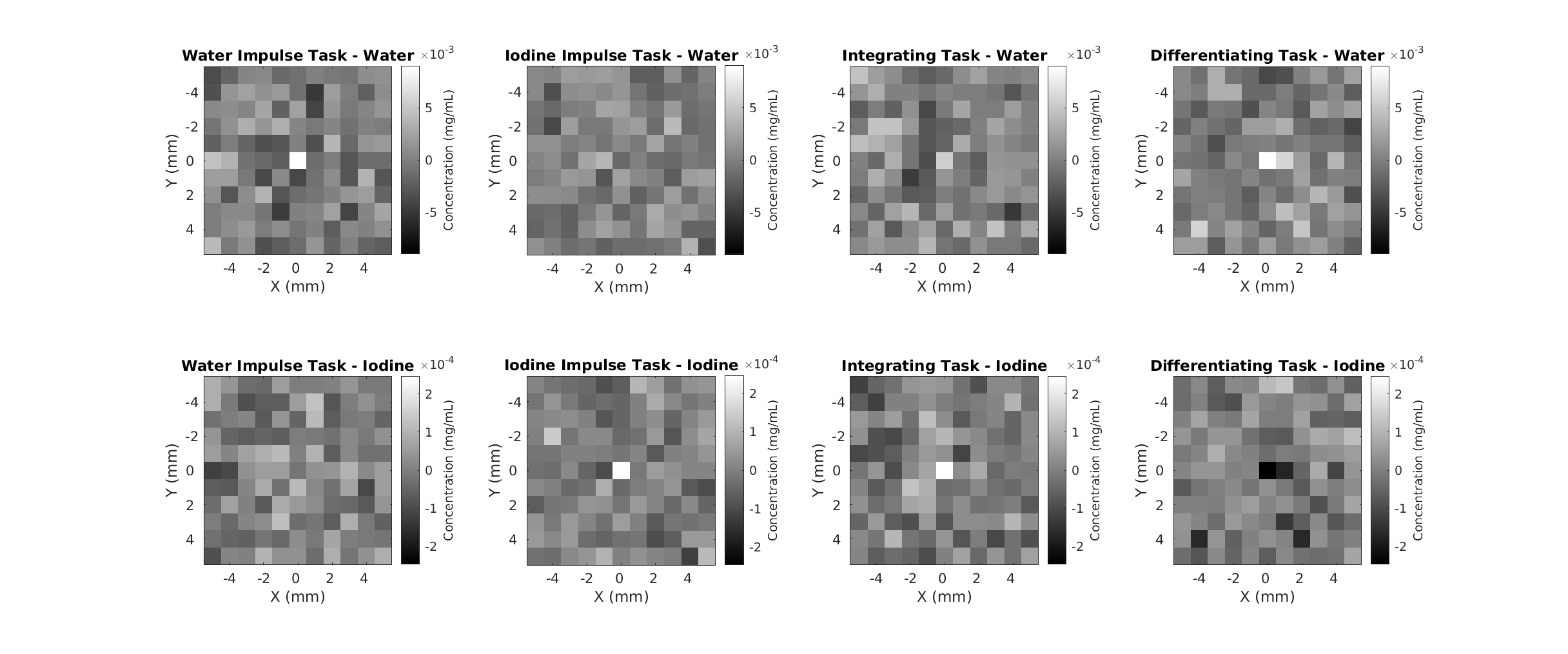}
    \vspace{-13mm}
    \caption{Reconstructed impulse stimuli in noise. Note that noise is anti-correlated between the water and iodine channels. Detectability index for the water task is 14.0, for the iodine task it is 14.0, for the integrating task (both impulses added) it is 27.7 and for the differentiating task (impulses subtracted) 4.8. Which results in a separability index of 0.174.}
    \vspace{-7mm}
    \label{fig:peturbationResponse}
\end{figure*}
% NOTE: Performance Metric for Design Optimization?

In the following sections, we will make comparisons between varied spectral CT system designs including various SSF configurations. To make a quantitative comparisons between those systems, we employ a performance evaluation framework for material decomposition using spectral CT. We introduce a new metric of separability as part of this framework. Toward this end, we write local approximations of the transfer function and noise for the spectral data acquisition and reconstruction.

\subsubsection{Transfer Function} We begin with the assumption that the relationship between the estimated material density images, $\mathbf{\hat{x}}$ and the ground truth, $\mathbf{x}$, is approximately linear for small changes to the ground truth as shown in the expression below:
\begin{gather}
    \mathbf{\Delta \hat{x}} = \Big[\frac{\boldsymbol{\partial}\mathbf{\hat{x}}}{\boldsymbol{\partial}\mathbf{x}}\Big] \mathbf{\Delta x} = \Big[\frac{\boldsymbol{\partial}\mathbf{\hat{x}}}{\boldsymbol{\partial}\mathbf{\bar{y}}}\Big] \Big[\frac{\boldsymbol{\partial}\mathbf{\bar{y}}}{\boldsymbol{\partial}\mathbf{x}}\Big] \mathbf{\Delta x}
\end{gather}
Here we are using the following notation for the Jacobian matrix between two vectors
\begin{gather}
    \Big[\frac{\boldsymbol{\partial}\mathbf{u}}{\boldsymbol{\partial}\mathbf{v}}\Big]_{i,j} = \frac{\partial u_i}{\partial v_j}.
\end{gather}

For example, the Jacobian matrix relating the measurement model, $\mathbf{\bar{y}}$, to the ground truth, $\mathbf{x}$ is
%
%%% WEB - please check that you like the above
% 
%% MTT - resolved. looks great

\begin{equation}
    \Big[\frac{\boldsymbol{\partial}\mathbf{\bar{y}}}{\boldsymbol{\partial}\mathbf{x}}\Big] = -\mathbf{S} \mathbf{D_x} \mathbf{Q} \mathbf{A}
    \label{eq:dYdX}
\end{equation}
where $\mathbf{D_x} = D\{\exp(-\mathbf{Q}\mathbf{A}\mathbf{x})\}$ are diagonal weights that depend on the object $\mathbf{x}$. 

%Since we are interested in the image properties of converged estimates, $\mathbf{D_x}$ is evaluated at the solution, $\mathbf{\hat{x}}$. 

%Note that for the purposes of image property prediction, the weights $\mathbf{D_x}$ can typically be effectively approximated directly from the measured data without circular dependence on knowledge of $\mathbf{\hat{x}}$.

Following the differentiation method described in \cite{fessler1996spatial},
\begin{gather} 
    \Big[\frac{\boldsymbol{\partial}\mathbf{\hat{x}}}{\boldsymbol{\partial}\mathbf{\bar{y}}}\Big]  = -\Big[\frac{\boldsymbol{\partial}^2\Phi}{\boldsymbol{\partial}\mathbf{\mathbf{x}}^2}\Big]^{-1} \Big[\frac{\boldsymbol{\partial}^2\Phi}{\boldsymbol{\partial}\mathbf{\mathbf{x}}\boldsymbol{\partial}\mathbf{\mathbf{\bar{y}}}}\Big] \\
    \Big[\frac{\boldsymbol{\partial}\mathbf{\hat{x}}}{\boldsymbol{\partial}\mathbf{\bar{y}}}\Big] = -\Big[\mathbf{F} + \mathbf{R}\Big]^{-1} \Big[\mathbf{A}^T \mathbf{Q}^T \mathbf{D_x}^T \mathbf{S}^T \mathbf{\Sigma_y^{-1}} \Big]
    \label{eq:dXdY}\\
    \mathbf{F} = \mathbf{A}^T \mathbf{Q}^T \mathbf{D_x}^T \mathbf{S}^T \mathbf{\Sigma_y^{-1}} \mathbf{S} \mathbf{D_x} \mathbf{Q} \mathbf{A}
\end{gather}
where $\mathbf{F}$ is the Fisher information matrix. This is the Hessian of the objective function data fidelity term. It describes the precision of the multi-material estimates based on the information contained in the measured data. Because the estimates include multiple materials, $\mathbf{F}$ contains information about cross-material as well as cross-voxel correlations.

As described in \cite{wang2019local}, we can combine \eqref{eq:dYdX} and \eqref{eq:dXdY} to arrive the following expression for a local transfer function
\begin{equation}
    \Big[\frac{\boldsymbol{\partial}\mathbf{\hat{x}}}{\boldsymbol{\partial}\mathbf{x}}\Big] = \Big[\frac{\boldsymbol{\partial}\mathbf{\hat{x}}}{\boldsymbol{\partial}\mathbf{\bar{y}}}\Big] \Big[\frac{\boldsymbol{\partial}\mathbf{\bar{y}}}{\boldsymbol{\partial}\mathbf{x}}\Big] = \Big[\mathbf{F} + \mathbf{R}\Big]^{-1} \mathbf{F}
    \label{eq:signalTransfer}
\end{equation}

\vspace{-2mm}

Here we can see that the transfer function depends on the relationship between the Fisher information, $\mathbf{F}$, and the regularization, $\mathbf{R}$. This expression characterizes the spatial blur (e.g. local impulse response) as well as the cross-talk between materials. 
These biases are introduced and controlled by the regularization term.

\subsubsection{Covariance}

Approximations for the covariance including cross-material correlations was defined in \cite{wang2020prospective} as follows
\begin{equation}
    \mathbf{\Sigma_x} = \Big[\frac{\boldsymbol{\partial}\mathbf{\hat{x}}}{\boldsymbol{\partial}\mathbf{\bar{y}}}\Big]^T \mathbf{\Sigma_y} \Big[\frac{\boldsymbol{\partial}\mathbf{\hat{x}}}{\boldsymbol{\partial}\mathbf{\bar{y}}}\Big] = \Big[\mathbf{F} + \mathbf{R}\Big]^{-1} \mathbf{F} \Big[\mathbf{F} + \mathbf{R}\Big]^{-1}
    \label{eq:cov}
\end{equation}
Spectral CT data with less noise, greater resolution, or greater separability will improve the precision, or the amount of information contained in $\mathbf{F}$, which will lower the variability in $\mathbf{\hat{x}}$. Higher regularization can decrease the noise but at the cost of a biased solution as described by \eqref{eq:signalTransfer}. 
Note that the Cramer-Rao lower bound is $\mathbf{\Sigma_x} \geq \mathbf{F}^{-1}$. From \eqref{eq:signalTransfer} and \eqref{eq:cov}, we can see that as the regularization approaches zero, the transfer function approaches identity and the covariance approaches the inverse of the Fisher information matrix. Assuming that $\mathbf{F}$ is invertible, this satisfies the condition for an efficient unbiased estimator, achieving equality to the Cramer-Rao lower bound.

\subsubsection{Detectability} We aim to apply these expressions for the signal transfer function and covariance to a single quality metric for the detectability of a signal, $\mathbf{w}$. For relatively small signals, where the linearity approximation holds, we can express the detectability index as
\begin{equation}
    d^2(\mathbf{w}) = \mathbf{w}^T \Big[\frac{\boldsymbol{\partial}\mathbf{\hat{x}}}{\boldsymbol{\partial}\mathbf{x}}\Big]^T  \mathbf{\Sigma_x}^{-1}   \Big[\frac{\boldsymbol{\partial}\mathbf{\hat{x}}}{\boldsymbol{\partial}\mathbf{x}}\Big] \mathbf{w}  = \mathbf{w}^T \mathbf{F} \mathbf{w}
\end{equation}
This formulation of detectability index performance of an ideal Bayesian observer with the task of discriminating the presence or absence of the known signal, $\mathbf{w}$. 

It is important to note that for spectral CT, detectability index does not tell the full story. For example, a CT system may have high detectability for a known iodine contrast signal, but this does not mean it would be effective at telling whether that signal came from iodine or soft tissue. Thus, there is a need for a quantitative metric for material separability.

% There are valid arguments against using an idealized observer model to describe image quality. For one, this metric does not depend on the regularization whatsoever. However, in this work, our primary concern is system design, so it is reasonable to maximize the task-weighted Fisher information which will depend on system geometry, $\mathbf{A}$, and spectral sensitivity $\mathbf{S}$.

\subsubsection{Separability}

% NOTE: Missing a transpose

% NOTE: Maybe an appendix for condition number version of the formula

%%% WEB - This is ok. Let's see what the reviewers say. (They may want more info still...)
% MTT - cool. I'm happy with the way that sounds
Spectral CT systems are capable of enhanced material discrimination with respect to conventional CT. To illustrate the importance of material separability, Figure \ref{fig:peturbationResponse} shows example (iodine-water) material density estimates which can be achieved with a dual-energy CT system. Note that the noise is highly anti-correlated between materials which is a consequence of poor material separability. The first three columns represent the following stimuli, respectively: 1) a positive water density impulse, 2) a positive iodine impulse, and 3) an impulse of both iodine and water. We note that even a single-energy CT system will yield a relatively high detectability index for these tasks since  there is no need to determine whether the measured signal is coming from an increased density of water versus iodine. The fourth column shows a stimulus that is a water impulse minus an iodine impulse (e.g. decreased iodine uptake with respect to some baseline). For a single-energy CT system, detection of this stimulus can be difficult to differentiate with the absence of any signal at all; whereas spectral CT should have a distinct advantage. There is need for a quantitative metric which summarizes this separability which distinguishes spectral CT systems. While performance generally increases with increased x-ray exposures, we would like to decouple overall exposure from system design. Therefore, we propose the following normalized formulation for the separability index:
\vspace{-2mm}
\begin{equation}
    s^2_{k_1,k_2} = \frac{(\mathbf{w}_{k_1} - \mathbf{w}_{k_2})^T \mathbf{F} (\mathbf{w}_{k_1} - \mathbf{w}_{k_2})}{(\mathbf{w}_{k_1} + \mathbf{w}_{k_2})^T \mathbf{F} (\mathbf{w}_{k_1} + \mathbf{w}_{k_2})} 
    \label{eq:sep1}
\end{equation}

\vspace{-1mm}

\noindent where $\mathbf{w}_{k_1}$ and $\mathbf{w}_{k_2}$ are identical spatial signals corresponding to different materials. Moreover, we scale the stimuli such that $d^2(\mathbf{w}_{k_1}) = d^2(\mathbf{w}_{k_2})$. This metric can be interpreted as a ratio of the detectability of the material-differentiating task and the material-integrating task.

\vspace{-4mm}

\begin{equation}
    s^2_{k_1,k_2} = \frac{d^2(\mathbf{w}_{k_1} - \mathbf{w}_{k_2})}{d^2(\mathbf{w}_{k_1} + \mathbf{w}_{k_2})} 
    \label{eq:sep2}
\end{equation}

\vspace{-1mm}

% Assuming $\mathbf{w}_{k_2}\mathbf{F}\mathbf{w}_{k_1} \geq 0$, the separability index will satisfy $s_{k_1,k_2} \in (0,1)$, where $s_{k_1,k_2} = 0$ is the degenerate case where the material estimates are fully-correlated (e.g. single-energy CT) and where $s_{k_1,k_2} = 1$ for uncorrelated estimates. 
%%% WEB - don't understand the statement on units.... Need this sentence?
% MTT - Resolved. removed sentence.
% The weighting guarantees that the metric is unaffected by the units of any of the estimates and it normalizes the signals across different materials according to detectability.

%%% WEB - We may generalize the above to multiple materials
%%% You need a sentence or two of how this generalization happens, it isn't clear where this all comes from....  Perhaps show how this reduces to the 2D case?

%% MTT - Resolved. added a few extra steps in the derivation rather than jumping directly to the generalized formula. 
%% MTT - Also added a list of mathematical properties of separability. This is based on what Emil Sidky said in our spectral zoom meeting. Its THEORETICALLY possible to do a water/fat decomposition. But this separability metric will tell you why thats not a good idea.

The definition given in \eqref{eq:sep1} and \eqref{eq:sep2} applies to a two-material imaging case but there is a need for a generalized definition which can apply to the case of imaging three or more materials. For this generalization of separability, we define a new cross material matrix $\mathbf{C}$ which is constructed from $\mathbf{F}$ and the single-material task functions $\mathbf{w_{k_n}}$ (with the same spatial distribution) as shown below

\vspace{-3mm}

\begin{equation}
     C_{k_n,k_m} = \frac{\mathbf{w}_{k_n}^T \mathbf{F} \mathbf{w}_{k_m}}{\sqrt{\mathbf{w}_{k_n}^T \mathbf{F} \mathbf{w}_{k_m}}\sqrt{\mathbf{w}_{k_n}^T \mathbf{F} \mathbf{w}_{k_m}}}
     \label{eq:sep3}
\end{equation}

\vspace{-2mm}

The ratio in \eqref{eq:sep3} is a normalization which mirrors \eqref{eq:sep1}. As a result, the diagonal elements of $\mathbf{C}$ will be 1 and the off-diagonal elements will be between 0 and 1. Thus, the two-material separability may be written as

\vspace{-2mm}

\begin{equation}
    s^2 = \frac{[1, \enspace -1] \mathbf{C} [1, \enspace -1]^T}{[1,\enspace 1]\mathbf{C}[1,\enspace 1]^T} 
    \label{eq:sep4}
\end{equation}

\vspace{-1mm}

Because on the form of $\mathbf{C}$, \eqref{eq:sep4} is equivalent to the ratio of the smaller to the larger eigenvalues of the $2\times2$ matrix. For three or more materials, we extend the definition of separability as the inverse of the condition number of $\mathbf{C}$.

\vspace{-5mm}

\begin{gather}
    s^2 = \frac{1}{\text{cond} (\mathbf{C} )  } 
\end{gather}

\vspace{-1mm}
\noindent This definition is compatible with three or more materials for cases such as bone, water, and iodine decompositions, and multi-contrast imaging. %Yet, it reduces to the same definition of separability for the two material case for the two-material case.

This definition allows quantification of separability and has several desirable properties. Given the following assumptions:
\begin{enumerate}
    \item $\mathbf{w_{k_n}}\mathbf{F}\mathbf{w_{k_m}} >= 0 \quad \forall \enspace n,m$
    \item $\mathbf{w_{k_n}}\mathbf{F}\mathbf{w_{k_n}} = \mathbf{w_{k_m}}\mathbf{F}\mathbf{w_{k_m}} \quad \forall \enspace n,m$
\end{enumerate}
\noindent the separability index can be shown to have the following properties:
\begin{enumerate}
    \item Separability is bounded by [0, 1]
    \item Separability is unitless
    \item Separability is independent of overall scale of $\mathbf{F}$
    \item Separability is independent of proportional scaling of the task functions, $\mathbf{w_{k_1}},...,\mathbf{w_{k_N}}$
    \item Separability of a set of materials is less than or equal to the separability of any subset of those materials
\end{enumerate}

Single-energy CT systems will generally have $s\approx0$. There will typically be some small amount of information contained in the data due to beam hardening effects which change the sensitivity spectra, but it is generally considered insufficient for material decomposition. The case $s=1$ would mean the material density information is immediately available from the measurements. This is not possible with standard tomographic system designs, in part, due to the large overlap between mass attenuation spectra of physical materials. 

The separability index is a quantitative metric for how well conditioned the material decomposition portion of the problem will be. Independence on the scale of $\mathbf{F}$ means that separability does not change based on the x-ray source power or the overall dose delivered to the patient. It is not defined relative to the overall noise level. Rather, it describes to the sensitivity or conditioning of the material decomposition problem in a task-specific imaging context. It can be used to determine whether material decomposition is feasible for a specific spectral CT system and set of materials.

\vspace{-2mm}
\subsection{Design Optimization}

One of the advantages of SSFs relative to other technologies for spectral CT is the flexibility the design. In this section we present an experimental procedure for SSF design optimization through simulation and application of the separability index metric. 

We use the physical models presented in previous sections to model a cone-beam CT system with a 1100~mm source-to-detector distance, 830~mm source-to-axis distance, 0.556~mm pixel spacing, and 360 views per rotation. These geometric parameters are designed to match a configuration that can be achieved on a realistic physical system.% such as our bench-top cone-beam CT system. 

The SSF was positioned 380~mm from the source. The filter tile motion trajectory is generated by a velocity square-wave with a constant speed of 2~mm/s for 60 seconds (or 60 views) followed by the same speed in the opposite direction for another 60 seconds and then repeating. We fix the length of one period of filter tiles to 60~mm so that each 60 second interval covers two periods of the filter. Since the filter motion is linear, the proportion of the filter length period taken up by a material is the roughly the same as the proportion of views which are filtered by that material, for a certain fan angle. For example, if 20\% of the filter is gold, then roughly 20\% of the views will be gold-filtered for a given detector. 

The alternating linear motion of the filter results in an equal mapping between the relative duty-cycle of each filter tile material on the spatial layout of the SSF pattern and the duty-cycle in terms of exposure time corresponding to the same filter for any given detector pixel throughout the scan. That is, if a filter is composed of filter tiles with relative widths of 75\% erbium and 25\% tin, then linear filter motion will result in each detector is illuminated with an erbium-filtered beamlet for 75\% of views and a tin-filtered beamlet for 25\% of views. 

We parameterized the design of the SSF by the filter tile materials, filter tile width, and filter tile thickness. We limit our filter tile materials to the list shown in Table \ref{tab:filterMaterialsList} which are affordable and  accessible. 
We limited the maximum number of filter materials to four and we limited the minimum filter tile width to 5mm. We use simulations to find optimized designs for four tasks: water/iodine decomposition in small patients, water/iodine decomposition in large patients, water/iodine/gadolinium/gold decomposition in small patients, water/iodine/gadolinium/gold decomposition in large patients. 

\begin{table}[ht!]
\label{tab:filterMaterialsList}
\caption{List of possible filter materials and thicknesses}
\centering
\begin{tabular}{|c|c|c|}
\hline
100mm Cu & 250mm Cu & 500mm Cu \\ \hline
100mm Sn & 250mm Sn & 500mm Sn \\ \hline
100mm Pr & 250mm Pr & 500mm Pr \\ \hline
100mm Gd & 250mm Gd & 500mm Gd \\ \hline
100mm Er & 250mm Er & 500mm Er \\ \hline
100mm Lu & 250mm Lu & 500mm Lu \\ \hline
127mm Ta & 250mm Ta &  \\ \hline
127mm W  & 250mm W  &    \\ \hline
100mm Au & 250mm Au & 500mm Au \\ \hline
100mm Pb & 250mm Pb & 500mm Pb \\ \hline 
\end{tabular}
\vspace{.5mm}
\end{table}

Each of these four design optimizations were conducted using a specific numerical phantom. The phantom for water/iodine decomposition in a small patient consisted of a 160mm diameter cylindrical water tank centered on the axis of rotation. It contains seven 20mm cylindrical iodine inserts spaced evenly by polar angle and centered on the circle which is 60mm from the center of the tank. The inserts contain iodine at concentrations 5.00, 2.50, 1.00, 0.50, 0.25, 0.10, and 0.00 mg/mL.

\begin{figure*}[ht!]
    \centering
    \includegraphics[width=0.95\textwidth,trim=0 0mm 0 0mm, clip] {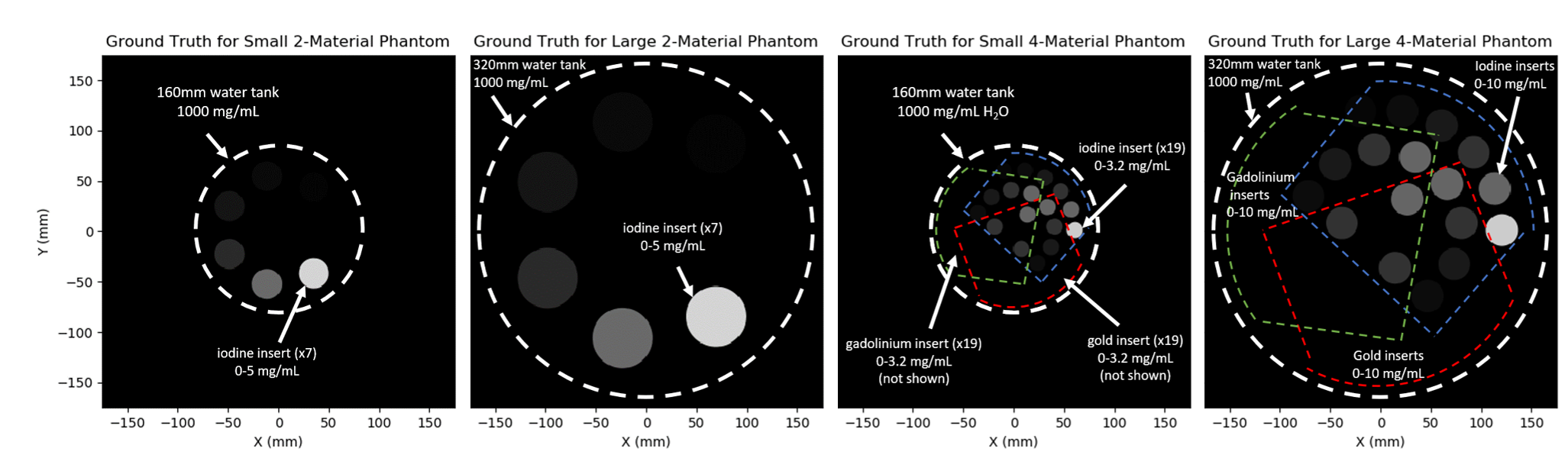}
    \vspace{-4mm}
    \caption{Ground truth digital phantoms for water/iodine imaging and water/iodine/gadolinium/gold imaging for small (160mm diameter) and large (320mm diameter) patient sizes.}
    \vspace{-6mm}
    \label{fig:phantom}
\end{figure*}

%%% WEB - why not move the phantom figures up here? Right?
%MTT - resolved.
The phantom for water/iodine decomposition in a large patient consisted of a 320mm diameter cylindrical water tank centered on the axis of rotation. The phantom also contains 20mm cylindrical inserts with even angular spacing and the same iodine concentrations as the previous phantom, but the inserts are centered on the circle which is 120mm in diameter. The phantom is represented in a voxelized multi-material image model with 2 materials (water and iodine), $350\times350$ voxels which are 1mm$\times$1mm. The dose absorbed by the patient, as described by \eqref{eq:dose}, was fixed to 1~mJ for the small patient and 5~mJ for the large patient. The transmissivity of the filters varies depending on the filter material and source spectra. For a thicker filter, the x-ray fluence must be higher to achieve the same patient dose with the filtered beamlets.

For optimized water/iodine/gadolinium/gold decomposition designs, we use the same 160mm and 320mm cylindrical water tanks for the small and large patient cases, respectively, but the inserts are 10mm in diameter for the small patient case and 20mm in diameter for the large patient case. The inserts contain mixtures of the three contrast materials. Each phantom contains an inner, middle, and outer ring of inserts. The outer ring contains single-contrast inserts, the middle ring contains two-material mixtures, and the inner ring contains three-material mixtures. The concentrations vary between 0-3.2 mg/mL. 

The metric for the design optimization is the 2-material or 4-material separability index using an impulse stimulus at the center of the object. Because this is a simulation study, we have access to the ground-truth Fisher information term, including the exact object-dependent weights. These are used to evaluate the predictive formula for separability index. The optimization is conducted by iteratively updating the estimated design parameters using a CMAES solver \cite{hansen2006cma}. 

After the designs have been optimized, we simulate noisy SSF projection data for the ideal design and run 1000 iterations of the MBMD reconstruction described in a previous section to demonstrate performance using particular designs. We included a quadratic smoothness penalty with cross-material weights optimized according to \cite{wang2020prospective}. The weights of this penalty term were set to $\approx 1$\% of the data fidelity term to keep bias to a minimum.

\section{Results}

\begin{figure*}[ht!]
    \centering
    \includegraphics[width=0.45\textwidth]{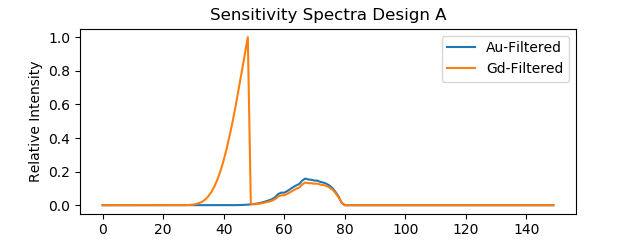}
    \includegraphics[width=0.45\textwidth]{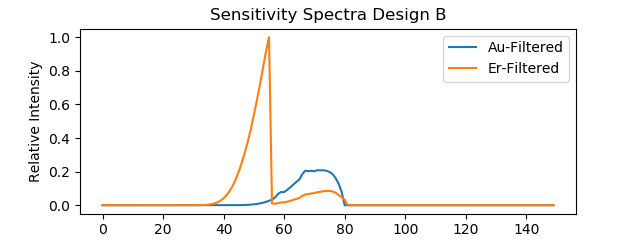}
    \includegraphics[width=0.45\textwidth]{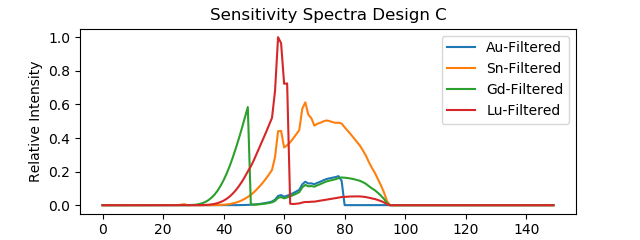}
    \includegraphics[width=0.45\textwidth]{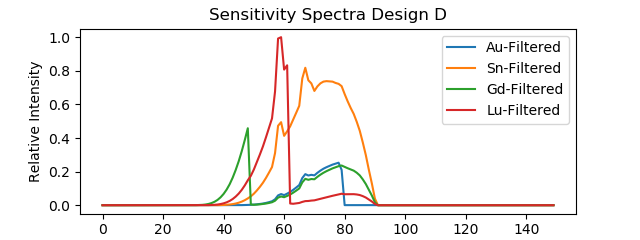}
    \vspace{-4mm}
    \caption{Filtered sensitivity spectra for the four SSF designs.}
    \label{fig:spectra}
    \vspace{-6mm}
\end{figure*}

\begin{figure}[h!]
    \centering
    \includegraphics[trim={0 0 0 0},clip,width=0.38\textwidth]{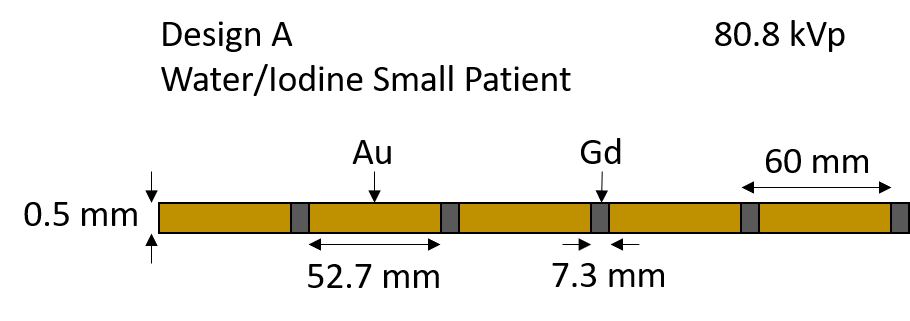}
    \includegraphics[width=0.38\textwidth]{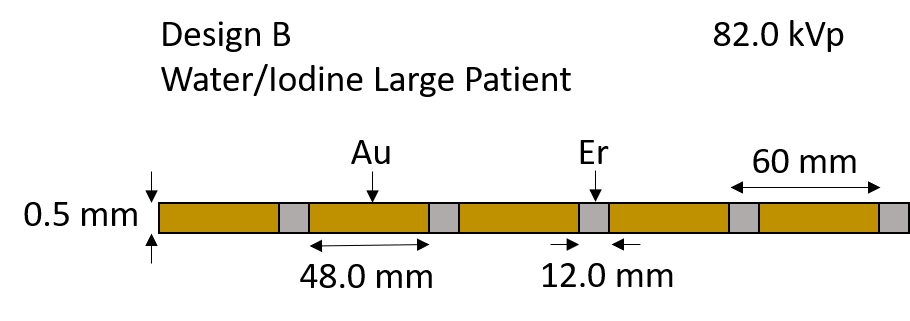}
    \includegraphics[width=0.38\textwidth]{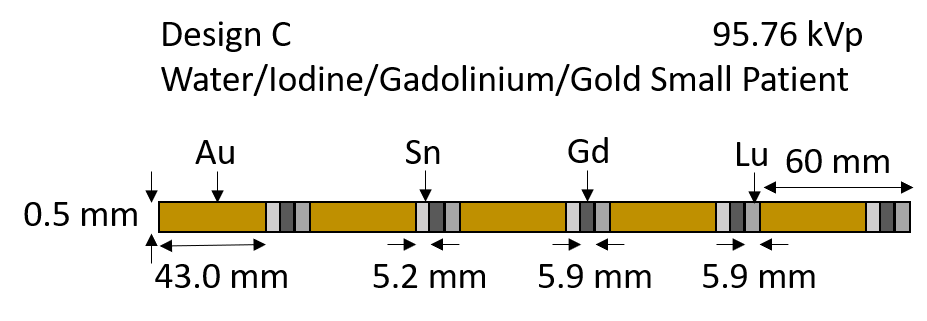}
    \includegraphics[width=0.38\textwidth]{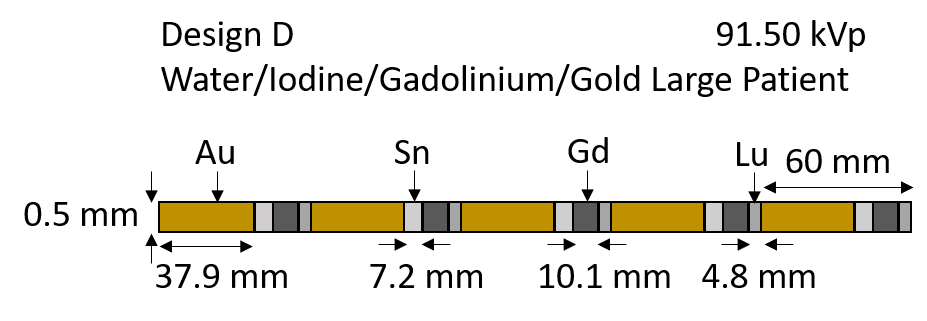}
    \vspace{-6mm}
    \caption{Optimized design results for the four experiments}
    \vspace{-7mm}
    \label{fig:ssf_design}
\end{figure}

The optimized SSF designs for each application case are shown in Figure \ref{fig:ssf_design}. For all cases, the optimized filter thickness was found to be 0.5mm which was the maximum allowed. This is not surprising because thicker filters lead to more filtration and more dissimilar spectra. The relative width of each filter in the optimized design is not equal. Gold, for example, is very wide in each of the designs. One possible explanation for this is that gold is very dense, so it attenuates more than some of the other materials, since the final results all have the maximum thickness of 0.5mm. These relative widths could be effectively normalizing the total number of photons delivered for each spectral channel.   

The optimized filter for water/iodine imaging in a small patient consisted of a repeating pattern of a 52.7mm wide gold filter tile and a 7.3mm wide gadolinium filter tile. The optimized source voltage for this case was 80.8kV. For water/iodine imaging in large patients, the optimized source voltage is 82.0kV and the optimized materials were gold and erbium with widths 48.0mm and 12.0mm, respectively. For water/iodine/gadolinium/gold imaging in both small and large patients, the optimized  filter was made of gold, tin, gadolinium. For small patients the widths were 43.0mm gold, 5.2mm tin, 5.9mm gadolinium, and 5.9mm lutetium. For large patients, the widths were 37.9mm gold, 7.2mm tin, 10.1mm gadolinium, and 4.8mm lutetium. The sensitivity spectra (including detector sensitivity) for each case are shown in Figure \ref{fig:spectra}.

\begin{figure*}[htb!]
    \centering
    \includegraphics[trim={0 0 0 8mm},clip,width=0.9\textwidth]{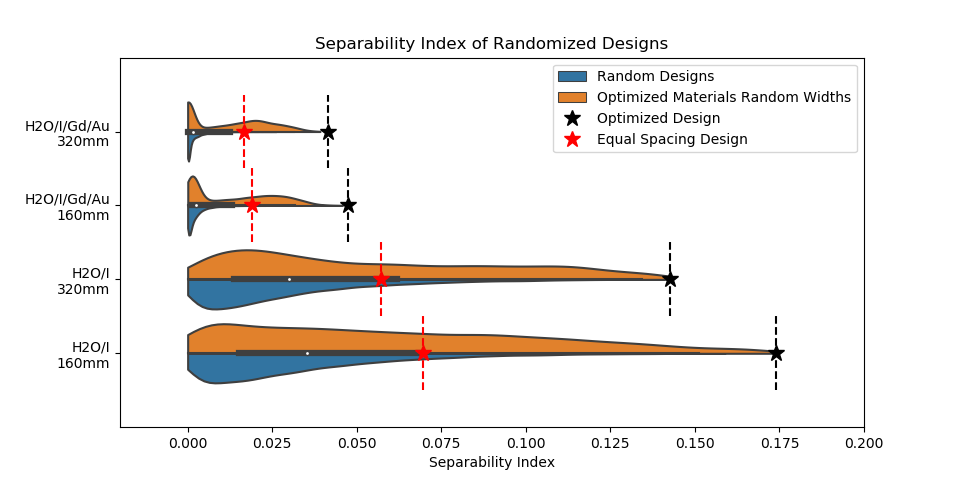}
    \vspace{-5mm}
    \caption{Violin plots showing the distribution of separability metrics for random designs.The final optimized designs are shown with a black star.}
    \vspace{-3mm}
    \label{fig:violin}
\end{figure*}

The violin plot in Figure \ref{fig:violin} serves to compare the separability index for optimized designs to randomized designs (uniformly distributed within design parameter ranges) to illustrate the importance of optimization. The orange distributions represent cases with the same filter materials and source voltage as the optimized case, but randomized relative filter tile widths. The blue distribution represents completely randomized designs including materials and source voltages. In both cases, the distribution of possible designs is centered much lower than the optimized case. The separability for the optimized design as well as the design using the optimized materials but with equal spacing are shown with black and red stars, respectively. 

\begin{figure*}[htb!]
\centering
\includegraphics[width=0.24\textwidth, trim= {10mm 3mm 15mm 4mm}, clip]{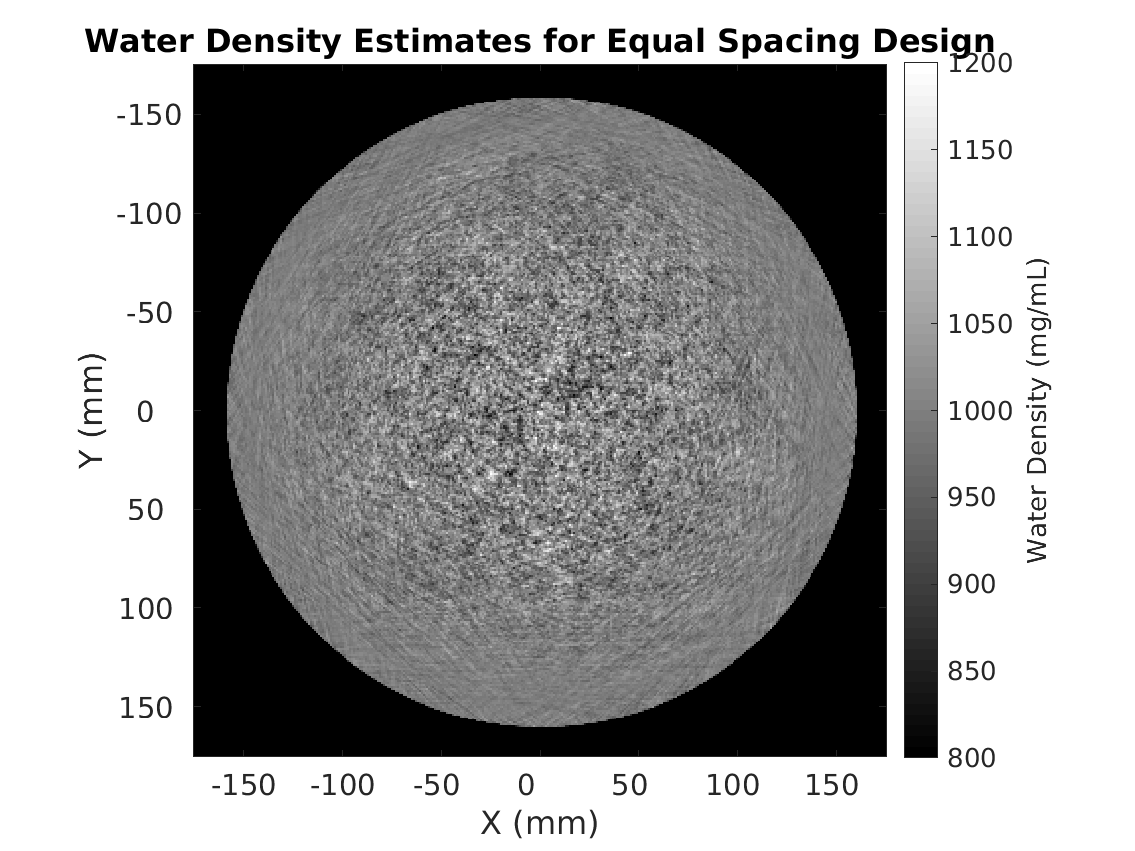}
\includegraphics[width=0.24\textwidth, trim= {10mm 3mm 15mm 4mm}, clip]{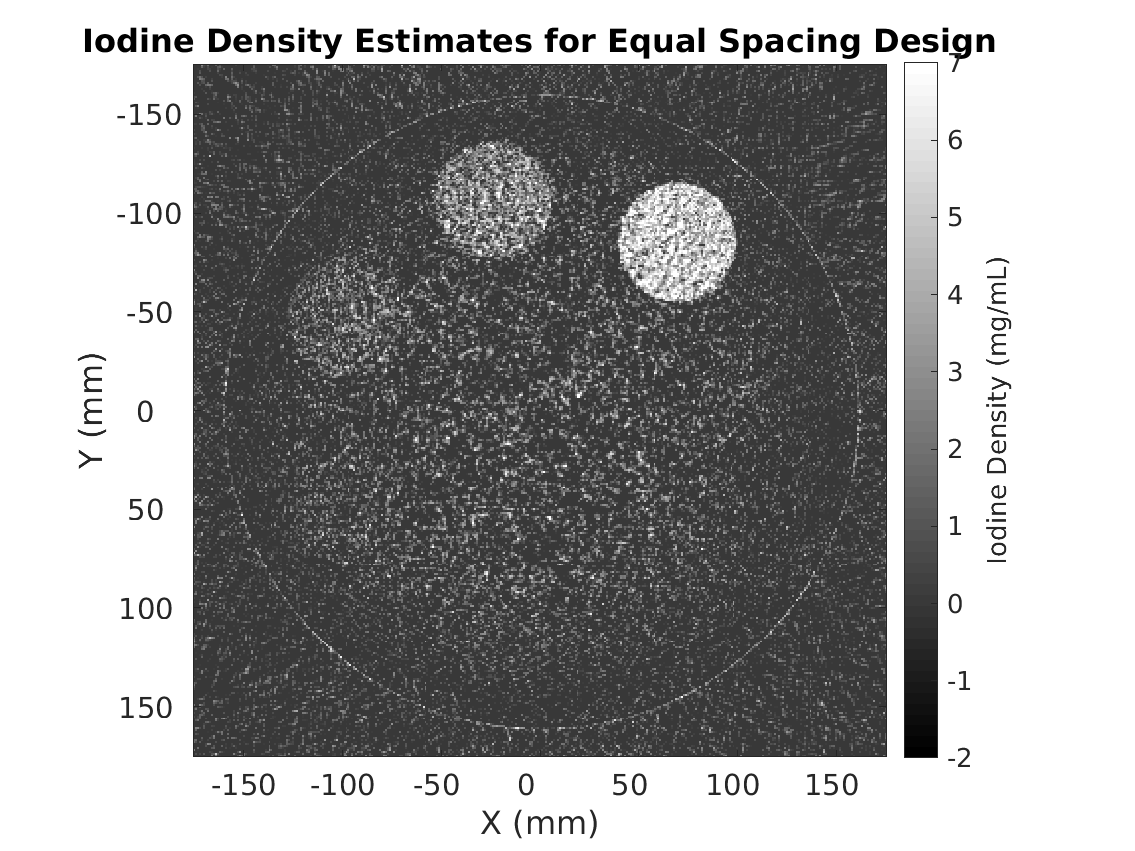}
\includegraphics[width=0.24\textwidth, trim= {10mm 3mm 15mm 4mm}, clip]{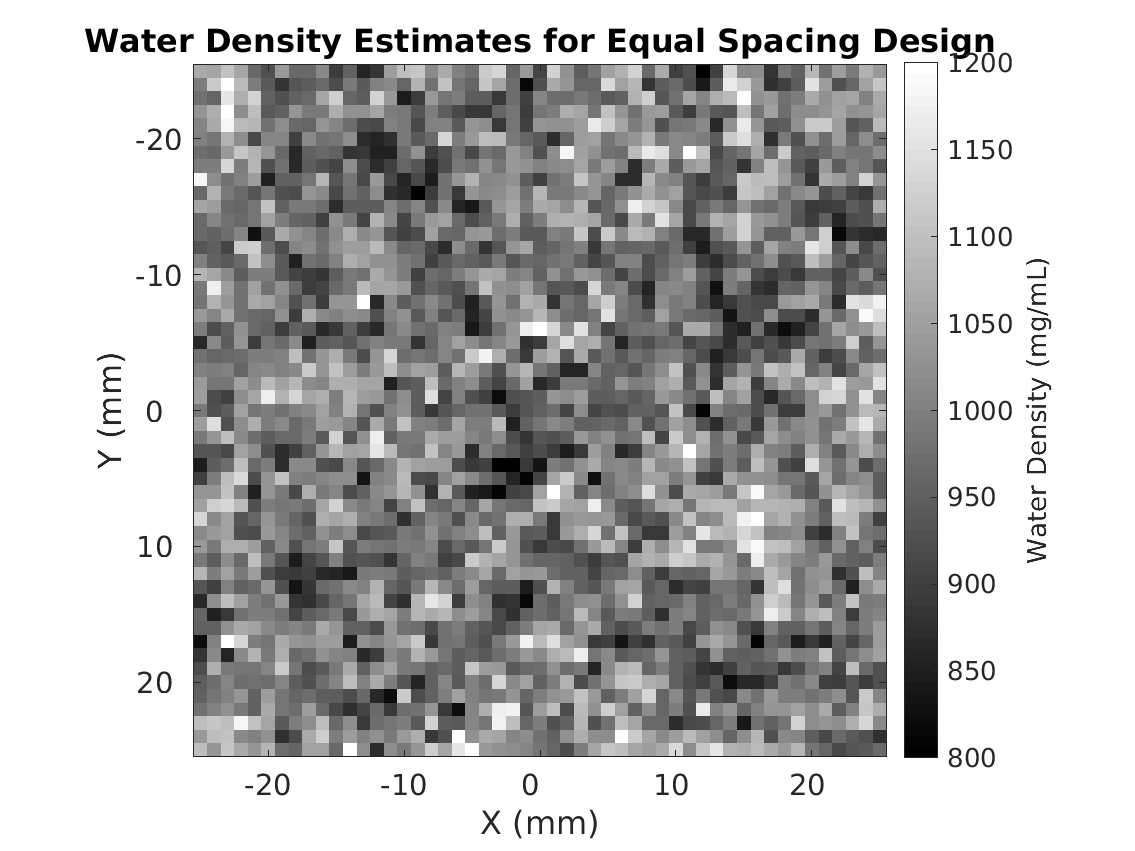}
\includegraphics[width=0.24\textwidth, trim= {10mm 3mm 15mm 4mm}, clip]{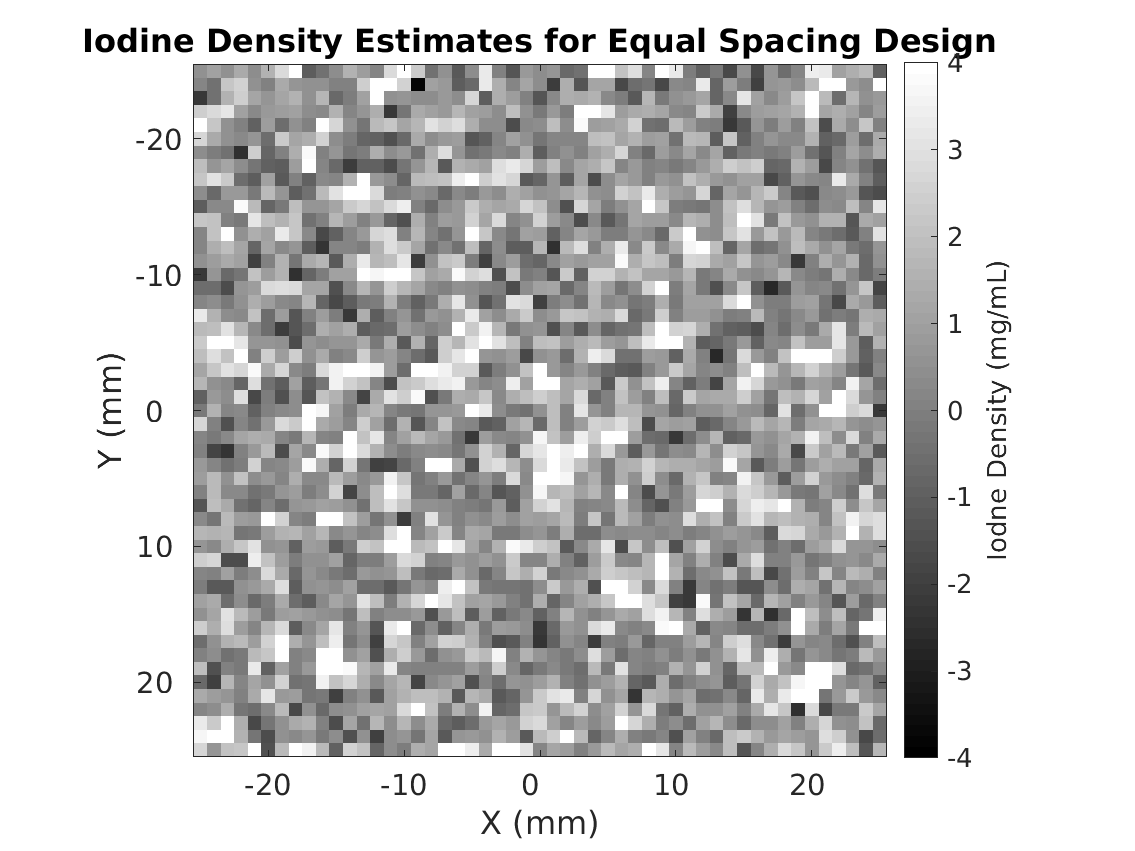}
\includegraphics[width=0.24\textwidth, trim= {10mm 3mm 15mm 4mm}, clip]{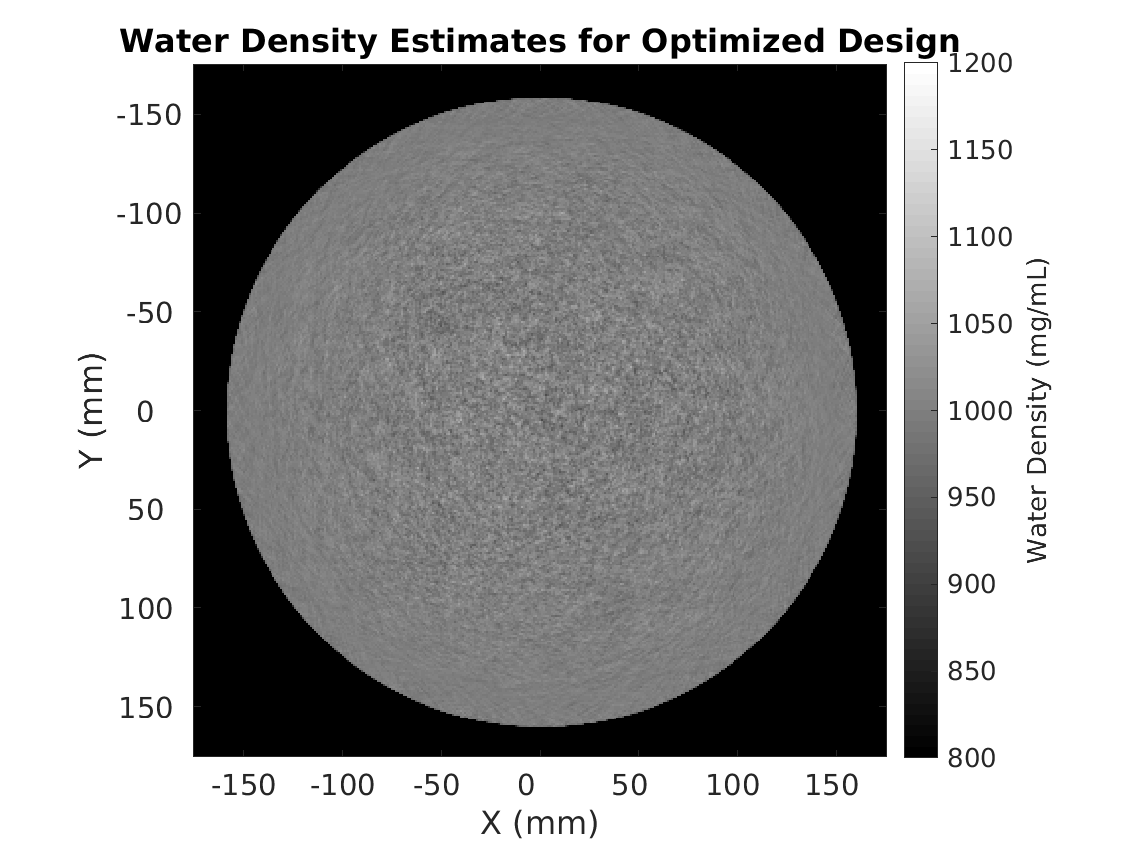}
\includegraphics[width=0.24\textwidth, trim= {10mm 3mm 15mm 4mm}, clip]{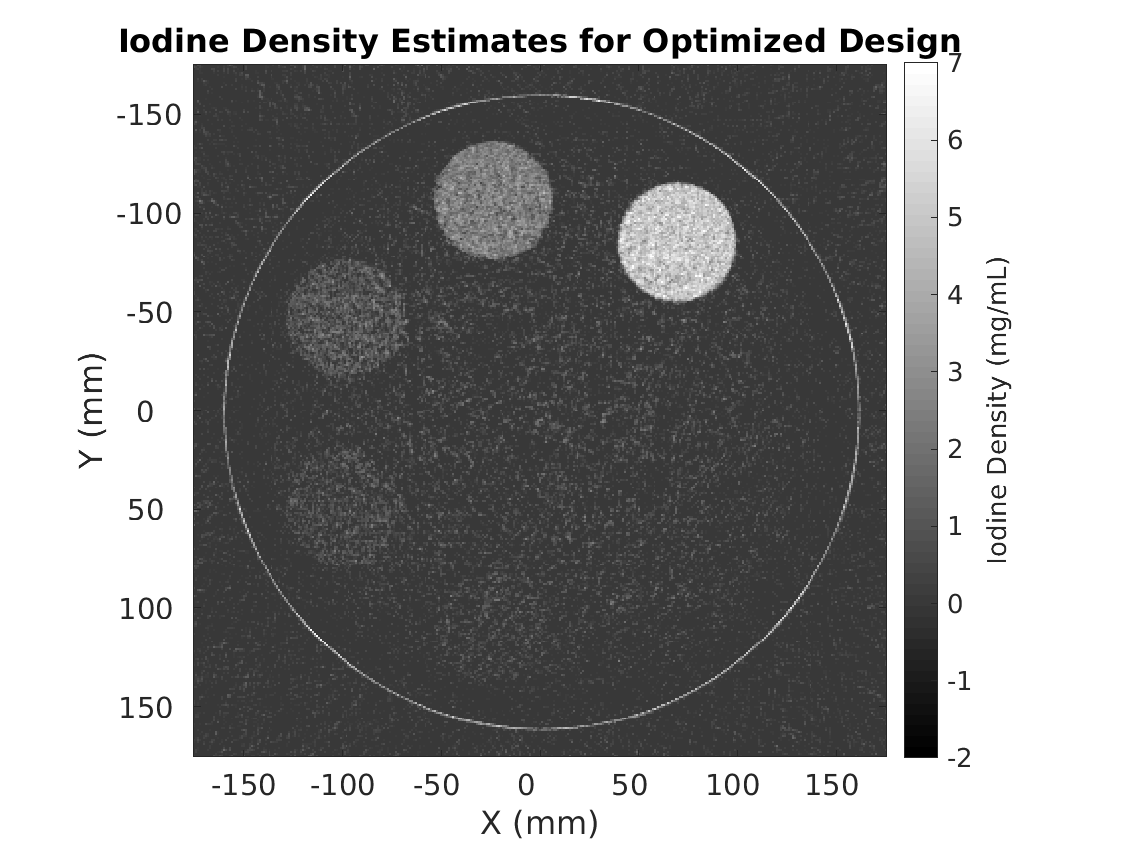}
\includegraphics[width=0.24\textwidth, trim= {10mm 3mm 15mm 4mm}, clip]{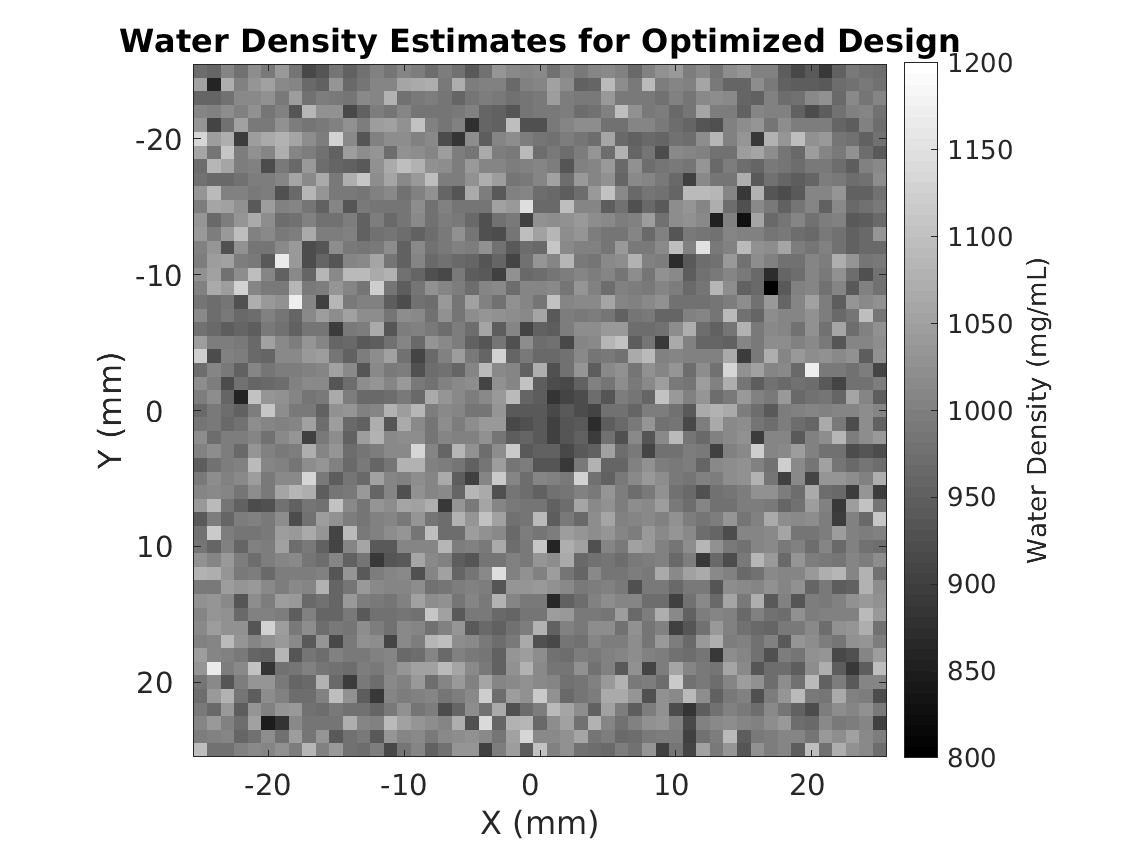}
\includegraphics[width=0.24\textwidth, trim= {10mm 3mm 15mm 4mm}, clip]{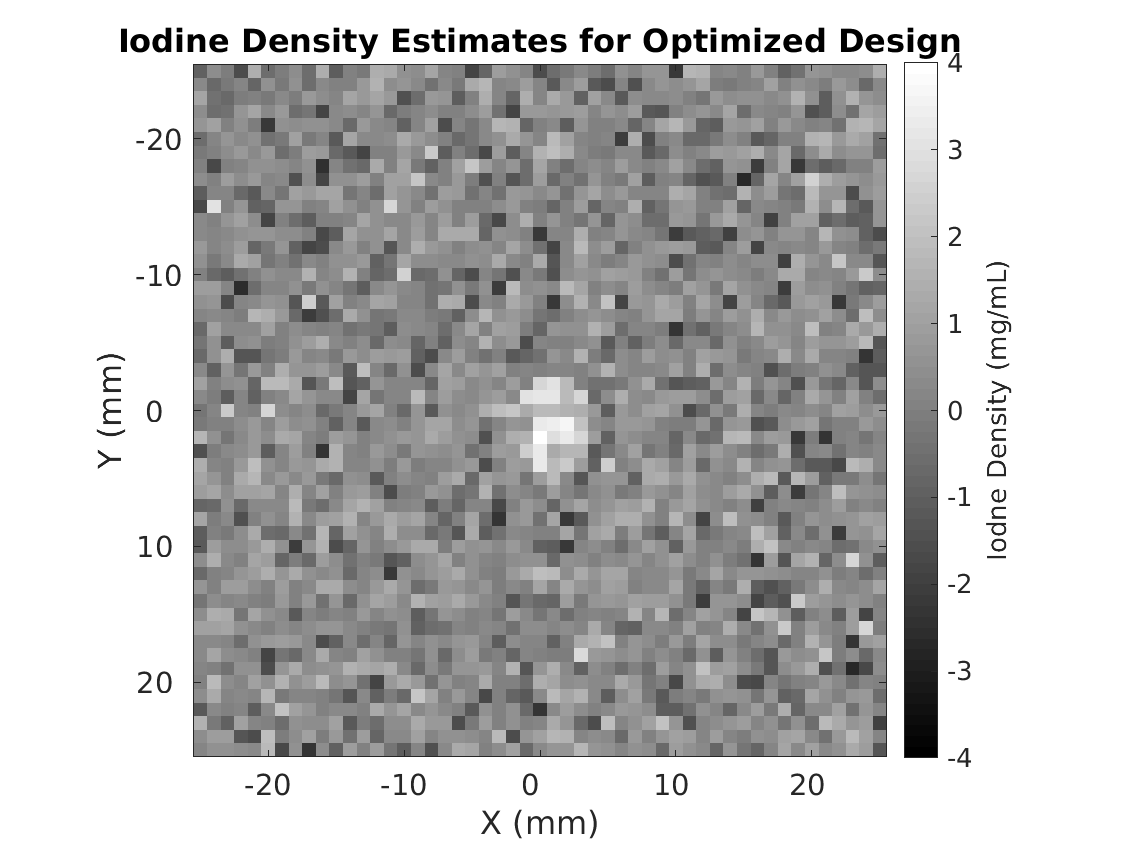}
\vspace{-3mm}
\caption{Material density estimates for phantom B. The top row corresponds to the equal-spacing design. The bottom row corresponds to the optimized design. The first two columns show the material density estimates. The second two columns show the perturbation response to a 7mm disc of -80mg/mL water and +2.0 mg/mL iodine to illustrate the improved separability. Severe noise correlations in the equal spacing design make it difficult to visualize the signal.}
\vspace{-4mm}
\label{fig:reconComparison}
\end{figure*}

Finally, for the large patient water/iodine imaging case, we show, in Figure \ref{fig:reconComparison}, a side-by-side comparison of the reconstructed iodine density estimates between the optimized design case, and the case using optimized material and source voltage but equal spacing of filter widths. The lower separability of the equal-spacing case leads to poor conditioning for the material separation part of the estimation problem. As a result the noise correlation between the two materials for the equal-spacing case is much larger. By inspecting the comparison in Figure \ref{fig:reconComparison} we an conclude that the noise in the iodine estimates is also much lower for the optimized design case. This illustrates the importance of a quantitative design optimization process using a predictive metric such as separability index.

% A one dimensional cross section of the optimization space is shown in Figure \ref{fig:optPlot} which plots separability index vs gold filter width. Note that the period of the filter v7  

\begin{figure}[htb!]
    \centering
    \includegraphics[width=0.24\textwidth,trim=8mm 3mm 15mm 7mm, clip] {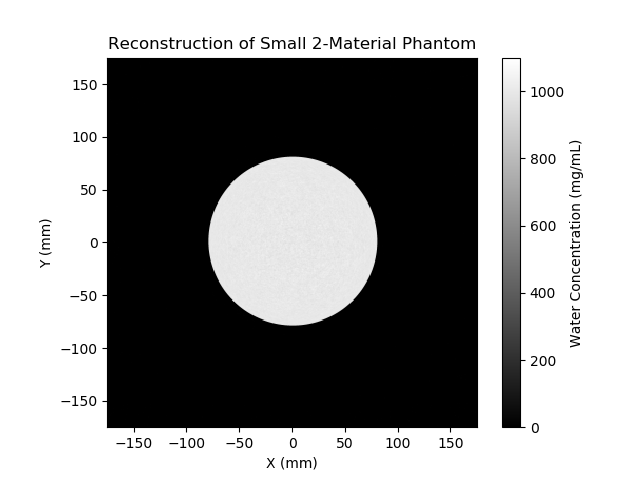}
    \includegraphics[width=0.24\textwidth,trim=8mm 3mm 15mm 7mm, clip ] {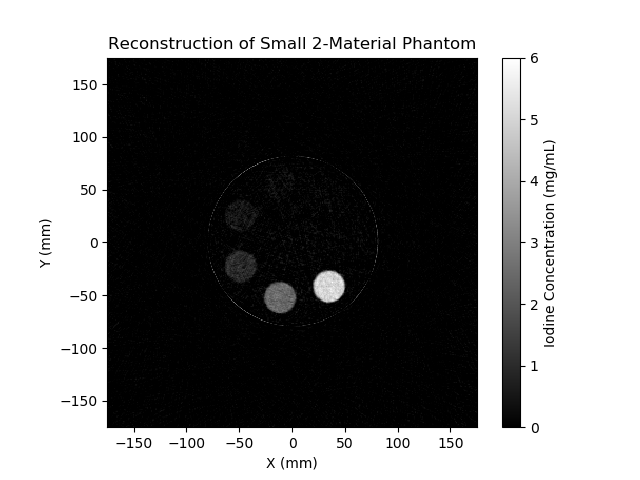}
    \caption{Small water/iodine phantom consisting of a 160mm diameter cylicindrical water tank with 30mm cylindrical iodine inserts.}
    \label{fig:recon_A}
\end{figure}

\begin{figure}[htb!]
    \centering
    \includegraphics[width=0.24\textwidth,trim=8mm 3mm 15mm 7mm, clip ] {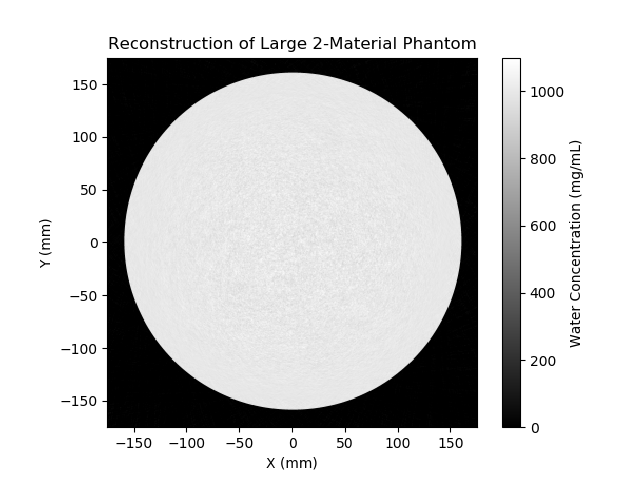}
    \includegraphics[width=0.24\textwidth,,trim=8mm 3mm 15mm 7mm, clip ] {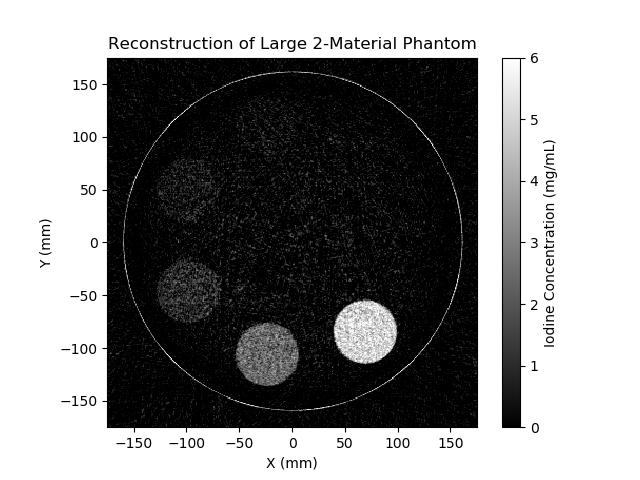}
    \caption{Large water/iodine phantom consisting of a 320mm diameter cylindrical water tank with 30mm cylindrical iodine inserts.}
    \label{fig:recon_B}
\end{figure}

\begin{figure*}[htb!]
    \centering
    \includegraphics[width=0.24\textwidth,trim=8mm 3mm 15mm 7mm, clip ] {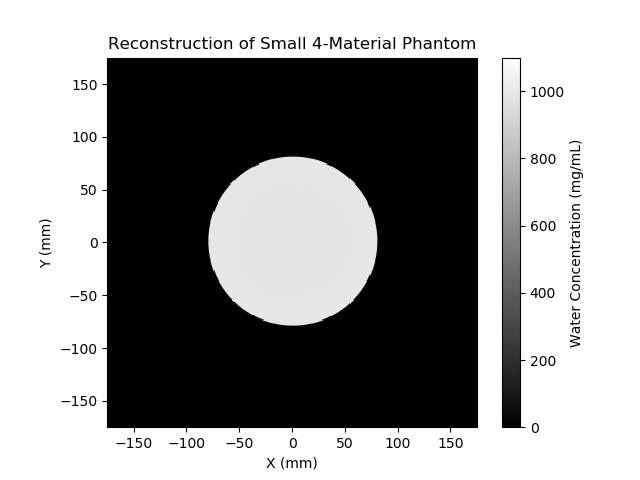}
    \includegraphics[width=0.24\textwidth,trim=8mm 3mm 15mm 7mm, clip ] {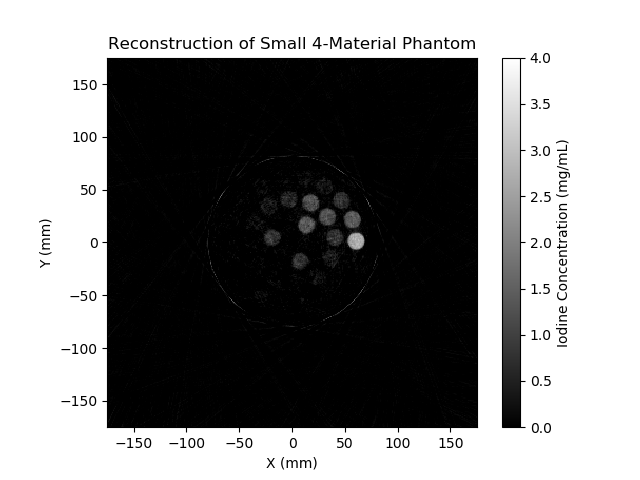}
    \includegraphics[width=0.24\textwidth,trim=8mm 3mm 15mm 7mm, clip ] {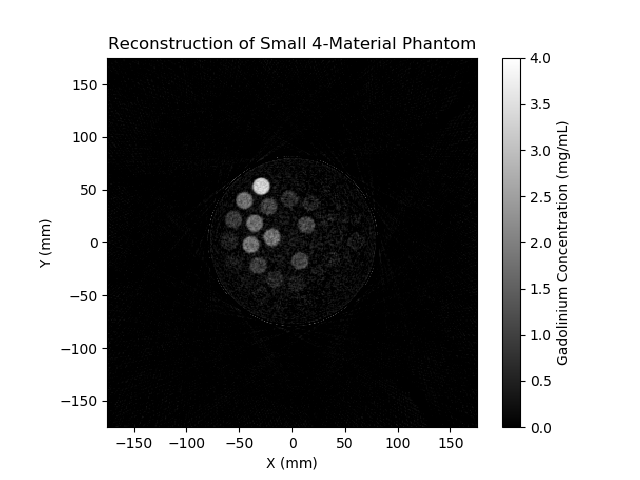}
    \includegraphics[width=0.24\textwidth,trim=8mm 3mm 15mm 7mm, clip ] {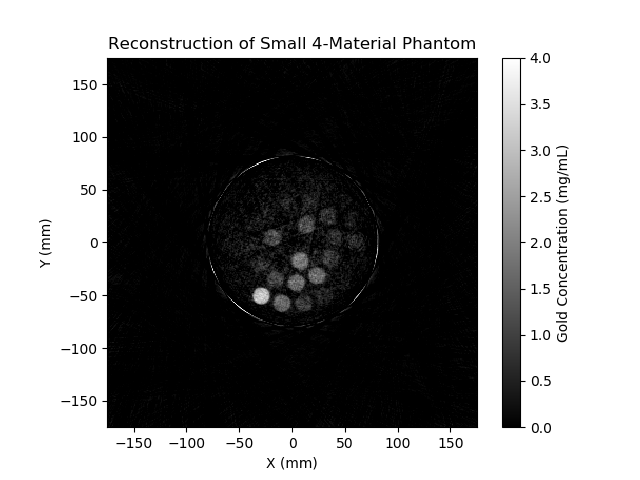}    
    \vspace{-4mm}
    \caption{Large water/iodine/gadolinium/gold phantom consisting of a 320mm diameter cylindrical water tank with 30mm cylindrical contrast inserts.}
    \label{fig:recon_C}
\end{figure*}

\begin{figure*}[htb!]
    \centering
    \includegraphics[width=0.24\textwidth,trim=8mm 3mm 15mm 7mm, clip ] {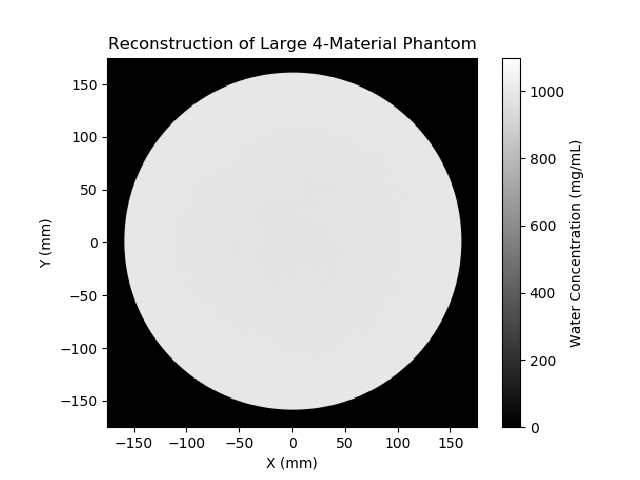}
    \includegraphics[width=0.24\textwidth,trim=8mm 3mm 15mm 7mm, clip ] {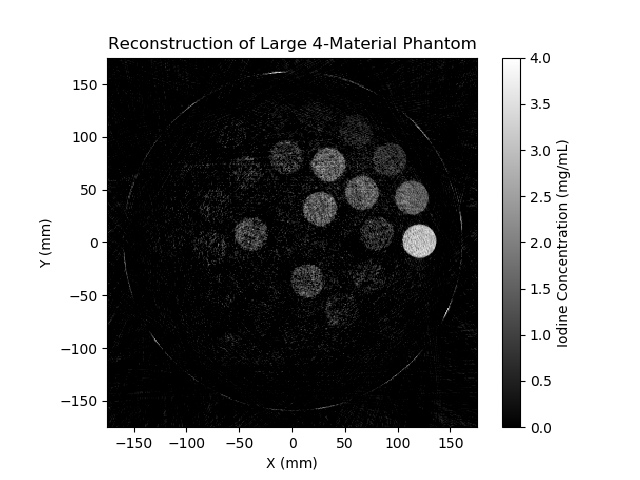}
    \includegraphics[width=0.24\textwidth,trim=8mm 3mm 15mm 7mm, clip ] {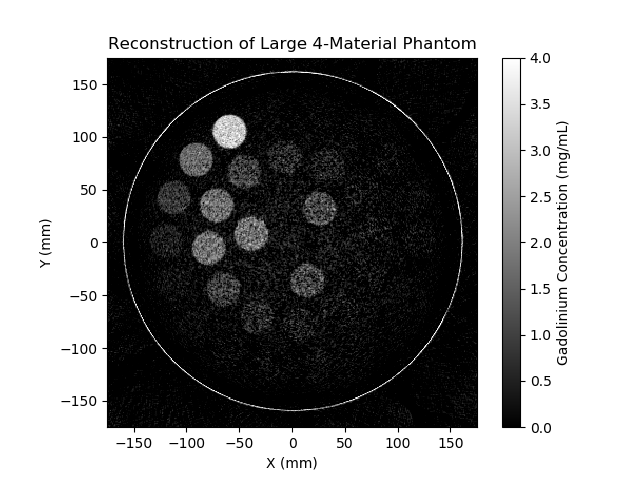}
    \includegraphics[width=0.24\textwidth,trim=8mm 3mm 15mm 7mm, clip ] {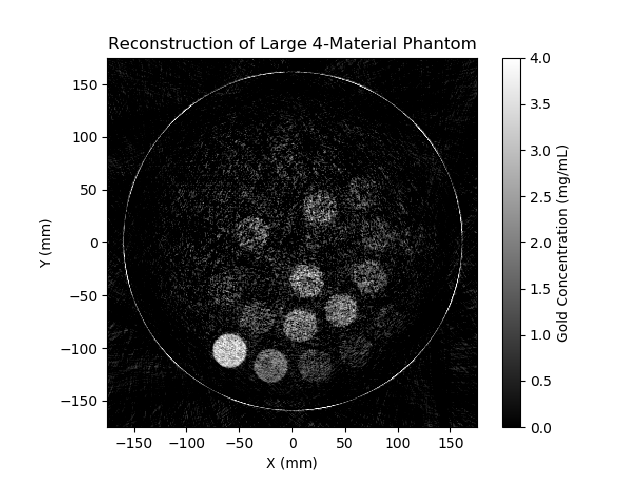}    
    \vspace{-4mm}
    \caption{Large water/iodine/gadolinium/gold phantom consisting of a 320mm diameter cylindrical water tank with 30mm cylindrical contrast inserts.}
    \vspace{-4mm}
    \label{fig:recon_D}
\end{figure*}

% \begin{figure*}[htb!]
%     \centering
%     \includegraphics[width=0.34\textwidth,trim= 8mm 3mm 15mm 7mm, clip] {figures/xi_B_1_compare.png}
%     \includegraphics[width=0.34\textwidth,trim= 8mm 3mm 15mm 7mm, clip ] {figures/xi_B_equalSpacing_1_compare.png}
%     \vspace{-4mm}
%     \caption{Comparison between optimized design and equal-spacing design for large-patient water/iodine imaging. }
%     \label{fig:reconComparison}
% \end{figure*}
% \begin{figure*}[htb!]
%     \centering
%     \includegraphics[width=\textwidth]{figures/optPlot.png}
%     \caption{Separability index vs gold filter width. This is a 1-dimensional cross section of the multi-dimensional design parameter space.}
%     \label{fig:optimization}
% \end{figure*}

\section{Conclusion}

Spectral CT with spatial-spectral filters is a promising new technology which enables enhanced material discrimination, quantitative material density estimation, and multi-material decomposition. They offer a method for adding spectral capability to existing CT systems without modifications to the source and with ordinary energy-integrating detectors.  In this work we have presented a generalized physical model for spectral CT and showed how it can be applied to model spatial-spectral filters and conduct a model-based material decomposition. We introduced a new image quality metric for spectral CT based on detectability to quantify separability. Finally, we presented a design optimization experiment, where the design parameters were tuned to maximize separability index for small and large patients, as well as two different material separation cases. 

Spatial-spectral filters have a flexible design framework which can be tuned based on the task of interest and the imaging target. Since most CT systems will be used for a variety of tasks, in future work, we intend to explore the possibility of adaptive CT with spatial-spectral filters (e.g. via different filter motions or two-dimensional filters). Additionally, the spatial-spectral filtering concept can be combined with other spectral techniques for improved separability. In the future, we are interested in using SSFs and kV-switching, or SSFs and energy-discriminating detectors in hybrid spectral CT systems to improve imaging performance, separability, and high-sensitivity quantitation.

\section*{Acknowledgments} 
This work was supported, in part, by NIH grant R21EB026849.

\bibliography{report}{}
\bibliographystyle{plain} % makes bibtex use spiebib.bst

\clearpage

\end{document}